\documentclass[conference]{IEEEtran}

\usepackage{paralist}
\usepackage{balance}
\usepackage{todonotes}
\usepackage{amsmath}
\usepackage{hyperref}
\usepackage{graphicx}
\usepackage{subfig}
\graphicspath{{fig/}}

\begin{document}
%\title{Revealing the Secrets of Artificial Neural Networks with Side-channel Analysis}
\title{CSI Neural Network: Using Side-channels to Recover Your Artificial Neural Network Information}

% author names and 
%\author{
%\IEEEauthorblockN{Lejla Batina}
%\IEEEauthorblockA{Institute for Computing and Information Sciences,
%	Radboud University,
%	Nijmegen, The Netherlands\\
%lejla@cs.ru.nl}
%\and
%\IEEEauthorblockN{Shivam Bhasin}
%\IEEEauthorblockA{Physical Analysis and Cryptographic Engineering, Temasek Laboratories at
%	Nanyang Technological University, Singapor \\  sbhasin@ntu.edu.sg}
%\and
%\IEEEauthorblockN{Dirmanto Jap}
%\IEEEauthorblockA{Physical Analysis and Cryptographic Engineering, Temasek Laboratories at
%	Nanyang Technological University, Singapor \\ djap@ntu.edu.sg}
%\and
%\IEEEauthorblockN{Stjepan Picek}
%\IEEEauthorblockA{Delft University of Technology, Delft, The Netherland
%s.picek@tudelft.nl}
%}


\author{\IEEEauthorblockN{Lejla Batina\IEEEauthorrefmark{1},
Shivam Bhasin\IEEEauthorrefmark{2},
Dirmanto Jap\IEEEauthorrefmark{2}, 
Stjepan Picek\IEEEauthorrefmark{3}}
\IEEEauthorblockA{\IEEEauthorrefmark{1}Institute for Computing and Information Sciences, Radboud University, Nijmegen, The Netherlands}
\IEEEauthorblockA{\IEEEauthorrefmark{2}Physical Analysis and Cryptographic Engineering, Temasek Laboratories at
	Nanyang Technological University, Singapore}
\IEEEauthorblockA{\IEEEauthorrefmark{3}Delft University of Technology, Delft, The Netherlands}}


\maketitle
\begin{abstract}
Machine learning has become mainstream across industries. Numerous examples proved the validity of it for security applications. In this work, we investigate how to reverse engineer a neural network by using only power side-channel information. To this end, we consider a multilayer perceptron as the machine learning architecture of choice and assume a non-invasive and eavesdropping attacker capable of measuring only passive side-channel leakages like power consumption, electromagnetic radiation, and reaction time. 

We conduct all experiments on real data and common neural net architectures in order to properly assess the applicability and extendability of those attacks. Practical results are shown on an ARM CORTEX-M3 microcontroller. Our experiments show that the side-channel attacker is capable of obtaining the following information: the activation functions used in the architecture, the number of layers and neurons in the layers, the number of output classes, and weights in the neural network. Thus, the attacker can effectively reverse engineer the network using side-channel information. 

Next, we show that once the attacker has the knowledge about the neural network architecture, he/she could also recover the inputs to the network with only a single-shot measurement. Finally, we discuss several mitigations one could use to thwart such attacks.
\end{abstract}

\section{Introduction}
\label{sec:introduction}

%what is the problem
%why is it interesting and important
%why is it hard
%what did other researcher do
%what are the key components of our approach

Machine learning, and more recently deep learning, has become hard to ignore for research in distinct areas, such as image recognition~\cite{Krizhevsky:2012:ICD:2999134.2999257}, robotics~\cite{KoberPM2012}, natural language processing~\cite{10.1007/978-3-642-14706-7_20}, and also security~\cite{Xu:2017:NNG:3133956.3134018, Kucera:2017:SPP:3133956.3134079} mainly due to its unquestionable practicality and effectiveness. 
Ever increasing computational capabilities of the computers of today and huge amounts of data available are resulting in much more complex machine learning architectures than it was envisioned before. As an example, AlexNet architecture consisting of 8 layers was the best performing algorithm in image classification ILSVRC2012~(\url{http://www.image-net.org/challenges/LSVRC/2012/}) classification task. In 2015, the best performing architecture for the same task was ResNet consisting of 152 layers~\cite{DBLP:journals/corr/HeZRS15}. This trend is not expected to stagnate any time soon, so it is prime time to consider machine learning from a novel perspective and in new use cases.

In this work, we focus on the widely used machine learning family of algorithms: the neural networks family. With the increasing number of design strategies and elements to use, fine tuning of hyperparameters of these algorithms is emerging as one of the main challenges. When considering distinct industries, we are witnessing an increase in intellectual property (IP) models strategies. Basically, for those cases when optimized networks are of commercial interest their details are kept undisclosed. For example, EMVCo (formed by MasterCard and Visa to manage specifications for payment systems and to facilitate worldwide interoperability) nowadays requires deep learning techniques for security evaluations~\cite{riscure}. This has an obvious consequence in: on one hand security labs generating (and using) neural networks for evaluation of products and on the other hand they treat them as IP, exclusively for their customers.

There are also other reasons for keeping the neural network architectures secret. Often, these pre-trained models might provide additional information regarding the training data, which can be very sensitive. For example, if the model is trained based on a medical record of a patient~\cite{Dowlin:2016:CAN:3045390.3045413}, confidential information could be encoded into the network during the training phase. Also, machine learning models that are used for guiding medical treatments are often based on a patient's genotype making this extremely sensitive from the privacy perspective~\cite{warfarin}.
Even if we disregard privacy issues, obtaining useful information from neural network architectures can lead to acquiring trade secrets from competition, which can lead to competitive products without violating intellectual property rights~\cite{Ateniese:2015:HSM:2829869.2829870}.
Hence, determining the layout of the network with trained weights is a desirable target for the attacker. 
One could ask the following question: why would an attacker want to reverse engineer the neural network architecture instead of just training the same network on its own? There are several reasons that are complicating this approach. First, the attacker might not have access to the same training set in order to train his own neural network. Second, as the architectures have become more complex, there are more parameters to tune and it could be extremely difficult for the attacker to pinpoint the same values for the parameters as in the architecture of interest.

%Q: social security number?

Our main question relates to the feasibility of reverse engineering such architectures.
Although binary analysis can already give useful information about the network, in practical cases, binary readback could be disabled by 
e.g., blocking JTAG access~\cite{khan2011embedded}. 
However, exploiting side-channel leakages remains a viable option.
Side-channel analyses have been widely studied in the community of information security and cryptanalysis, due to its potentially devastating impact on otherwise (theoretically) secure algorithms.  Concretely, it has been observed that different physical leakages from devices on which cryptography is implemented, such as timing delay, power consumption, and electromagnetic emanation (EM) during the computation of the data is dependent on the processed internal state and thus data. By statistically combining this physical leakage of the specific internal state and hypothesis on the data being manipulated, it is possible to recover the intermediate state being processed by the device. 

In this study, our aim is to highlight the potential vulnerabilities of standard (perhaps still naive from the security perspective) implementations of neural networks. 
At the same time, we are unaware of any neural network implementation in the public domain that includes side-channel protection. For this reason, we do not just point to the problem but also suggest some means of protection of neural networks against side-channel attacks.
Here, we consider some of the basic building blocks of neural networks: the number of hidden layers, the basic multiplication operation, and the activation functions. 
%One difference with side-channel attacks is that, the model used for the attack will be integer value based, whereas for neural networks, they often deal with floating-point number. Thus, the assumption of the leakage behavior or the leakage model have to be adjusted to approximate the behavior of the physical leakage.
Assuming that the multiplications are performed on one known and one unknown operand and by observing e.g., power consumption as the leakage, additional information about the output of the multiplication becomes available. In this case, different hypotheses of the possible values can be correlated with the leakage to recover the unknown input up to a certain precision. We show that for our target implementation, the value of an unknown input to the multiplication could be estimated with up to 0.01 precision. %Those results were possible by  choosing a suitable leakage model to emulate floating-point operations. Nevertheless, this is not a limiting factor and other models could be appropriate as well. 
%Also, when looking into the assumption on the adversary, there are other avenues that could be investigated. For example, the approach can be further strengthened if the attacker is allowed profiling a clone device and characterize all the leakages beforehand. This could prove as a good strategy in case of other platforms.
%This approach can be adopted to other platforms as well, by first profiling the device in order to obtain better approximation of the leakage model to mount the attack.

The complex structure of activation function often leads to conditional branching due to the necessary exponentiation and division operations.
Thus, conditional branching introduces input dependent timing differences resulting in different timing behavior for different activation function, allowing function identification. 
%In the experiments, we highlight that due to different mathematical operations used in activation, the timing delay information can be used to determine which activation function is used in the networks.
Basically, simply by observing the side-channel signatures, it is possible to deduce number of nodes, and also the number of layers in the networks.
%To recover the overall layout, by observing the timing difference, it is possible to deduce the number of nodes, and also number of layers in the networks. 
By using the usual divide-and-conquer approach for side-channel analysis, the information at each layer could be recovered, and the recovered information can be used as input for recovering the subsequent layers. Consequently, in this work, we show it is possible to recover the layout of unknown networks by exploiting the side-channel information.

To our best knowledge, this kind of observation has never been used before in this context. At least not for leveraging on (power/EM) side-channel leakages with reverse engineering the neural networks architecture as the main goal. We position our results in the following sections of this work.

The motivation for our work comes from ever more pervasive use of neural networks in security-critical applications and the fact that the architectures are becoming proprietary knowledge for the security evaluation industry. Thus, reverse engineering a neural net has become a new target for the adversaries and we need a better understanding of the vulnerability to side-channel leakages in those cases to be able to protect the users' rights and data.

\subsection{Related Work}

There are many papers considering machine learning and more recently, deep learning for improving the effectiveness of side-channel attacks. For instance, a number of works have compared the effectiveness of classical profiled side-channel attacks against various machine learning techniques~\cite{lerman2015template,jap2015support,picek2017side}.
Lately, several works explored the power of deep learning in the context of side-channel analysis~\cite{maghrebi2016breaking}. However, that line of work is putting a classifier from machine learning in the context of side-channel distinguishers i.e. the selection function leading typically to e.g., the key recovery.

On the other hand, using side-channel analysis in order to attack machine learning architectures has been much less investigated. % and to our best knowledge there exist only a few works taking this turn.
Shokri et al. investigate the leakage of sensitive information from machine learning models  about individual data records on which they were trained~\cite{7958568}. They show that such models are vulnerable to membership inference attacks and they also evaluate some mitigation strategies.
Song et al. show how a machine learning model from a malicious machine learning provider can be used to obtain information about the training set of a model~\cite{Song:2017:MLM:3133956.3134077}. 
Hua et al. were first to reverse engineer two convolutional neural networks, namely AlexNet and SqueezeNet through memory and timing side-channel leaks~\cite{Hua}. The authors measure side-channel through an artificially introduced hardware Trojan. They also need access to original training data for part of the attack, which might not always be available. Lastly, in order to obtain the weight of the neural networks, they attack very specific operation i.e., zero pruning~\cite{8192478}, which to an extent is more common for ReLU. 
%We consider their approach to be less realistic than ours since they need to insert a hardware Trojan to collect the memory traces of the FPGA accelerator where the experiments are conducted. They also experiment with obtaining the weights of neural networks but they succeed only by exploiting the zero pruning~\cite{8192478}. Such an approach could be realistic for specific settings but still is much less generic than ours. 
Wei et al. have also performed an attack on an FPGA-based convolutional neural network accelerator~\cite{DBLP:journals/corr/abs-1803-05847}.  They recovered the input image from the collected power consumption traces. The proposed attack exploits a specific design choice i.e., the line buffer in a convolution layer of a CNN.
In a nutshell, both previous reverse engineering efforts using side-channel information were performed on very special design choices for neural networks and having specific goals for the attacks.

%which allow optimized implementations.

Ohrimenko et al. used a secure implementation of MapReduce jobs and analyzed intermediate traffic between reducers and mappers~\cite{Ohrimenko:2015:OPL:2810103.2813695}. They showed how an adversary observing the runs of typical jobs can infer precise information about the inputs.
Xu et al. introduced controlled-channel attacks, which is a type of side-channel attack allowing an untrusted operating system to  extract  large  amounts  of  sensitive  information  from protected  applications~\cite{Xu:2015:CAD:2867539.2867677}.
Ohrimenko et al. discussed how machine learning algorithms data-oblivious algorithms can be exploited by various side-channels~\cite{Ohrimenko:2016:OMM:3241094.3241143}. Consequently, they propose data-oblivious machine learning algorithms  that prevent exploitation of side channels induced by memory, disk, and network accesses. Still, they note that side-channel attacks based on power and timing analysis are outside of the scope of their research.

Orthogonally to those works, we explore the problem of reverse engineering of neural networks from a more generic perspective and in a grey to black-box setting. To be specific, the closest previous works to ours have reverse engineered neural networks by using cache attacks which work on distinct CPUs and are basically micro-architectural attacks (although using timing side-channel). Our approach utilizes power side-channel on small embedded devices and it is supported by practical results obtained on a real-world architecture.

%is mounted on the hardware component executing the convolution in the first layer of a CNN. Their attacks is also special since they use the properties of the line buffer, which is an efficient hardware structure to implement convolutions. In this way they avoid the problem of finding the parameters of the architecture .

%\todo{emphasize the difference of our work when compared to other stuff}

\subsection{Contribution and Organization}
The main contributions of this paper are:
\begin{enumerate}
\item We describe a full reverse engineering of neural network parameters based on side-channel analysis. A combination of side-channel leakages is used to recover key parameters i.e., activation function, pre-trained weights, number of hidden layers and neurons in each layer. The proposed attack does not need any information on the (sensitive) training data as that information is often not even available to the attacker. We emphasize that, for our attack to work, we require only the knowledge of some inputs/outputs and side-channel measurements, which is a standard assumption for side-channel attacks. 
\item All the proposed attacks are practically implemented and demonstrated on two distinct microcontrollers (i.e. 8-bit AVR and 32-bit ARM), allowing full reverse engineering of the network architecture.
\item Further, a single trace input recovery attack has been proposed, which recovers a dataset when applied on the initial layers. This implies that the attacker can recover all the inputs tested with a known neural network, recovering each input from a single measurement. Such attacks can put user's sensitive data at great risk.
\item We highlight some interesting aspects of side-channel attacks when dealing with real numbers, unlike in everyday cryptography. For example, we show that even a side-channel attack that failed can provide sensitive information about the target due to precision error.
\item Finally, we propose a number of mitigation techniques that will render side-channel attacks more difficult.
\end{enumerate}
We emphasize again that the simplicity of our attack is its strongest point, as it minimizes the assumption on an adversary. This makes the underlying problem even more serious as the attack does not require any pre-processing, chosen-plaintext messages, etc.\\

The rest of this paper is organized as follows. In Section~\ref{sec:background}, we give details about specific machine learning algorithms we consider and side-channel analysis techniques we use. Section~\ref{sec:experiments} gives results on reverse engineering of various elements of neural networks and Section~\ref{sec:input_recovery} on input recovery attack. Section~\ref{sec:arm} demonstrates the feasibility of attack on modern 32-bit ARM microcontrollers. In Section~\ref{sec:countermeasures}, we briefly discuss possible countermeasures one could apply to make our attacks more difficult. Finally, in Section~\ref{sec:conclusion}, we conclude the paper and discuss potential future research directions.

\section{Background}
\label{sec:background}

In this section, we give details about artificial neural networks and their building blocks. Next, we discuss the concepts of side-channel analysis and several types of attacks we use in this paper.

\subsection{Artificial Neural Networks}

Artificial neural networks (ANNs) is an umbrella notion for all computer systems loosely inspired by biological neural networks. Such systems are able to ``learn'' from examples, which makes them a strong (and very popular) paradigm in the machine learning domain.
Any ANN is built from a number of nodes called artificial neurons. The nodes are connected in order to transmit a signal.
Usually, in an ANN, the signal at the connection between artificial neurons is a real number and the output of each neuron is calculated as a nonlinear function of the sum of its inputs. Neurons and connections have weights that are adjusted as the learning progresses. Those weights are used to increase or decrease the strength of a signal at a connection.
In the rest of this paper, we use the notions of an artificial neural network, neural network, and network interchangeably.

\begin{figure}
	\centering
	\includegraphics[width=0.8\linewidth]{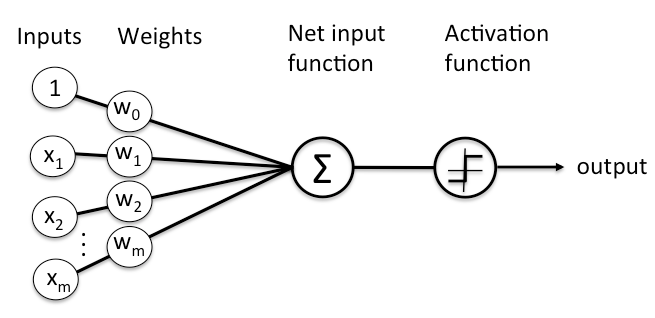}
	\caption{Depiction of an artificial neuron.}
	\label{fig:neuron}
\end{figure}

A very simple type of a neural network is called perceptron. A perceptron is a linear binary classifier applied to the feature vector. Each vector component has an associated weight $w_i$ and each perceptron has a threshold value $\theta$. The output of a perceptron equals ``1'' if the direct sum between the feature vector and the weight vector is larger than zero and ``-1'' otherwise.
A perceptron classifier works only for data that are linearly separable, i.e., if there is some hyperplane that separates all the positive points from all the negative points~\cite{Mitchell}.
We depict a model of an artificial neuron in Figure~\ref{fig:neuron}. In the case of the perceptron, the activation function is the step function.

By adding more layers to perceptron, we arrive to the multilayer perceptron algorithm. Multilayer perceptron (MLP) is a feed-forward neural network that maps sets of inputs onto sets of appropriate outputs. It consists of multiple layers of nodes in a directed graph, where each layer is fully connected to the next one. 
Consequently, each node in one layer connects with a certain weight $w$ to every node in the following layer.
Multilayer perceptron algorithm consists of at least three layers: one input layer, one output layer, and one hidden layer. Those layers must consist of nonlinearly activating nodes~\cite{Collobert}.

\begin{figure}
	\centering
	\includegraphics[width=0.9\linewidth]{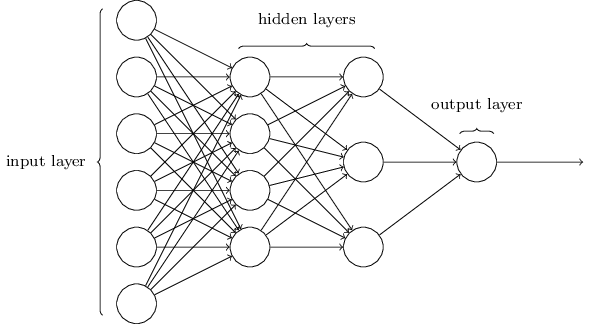}
	\caption{Multilayer perceptron.}
	\label{fig:mlp}
\end{figure}

We depict a model of a multilayer perceptron in Figure~\ref{fig:mlp}. Note, if there is more than one hidden layer, then it can be considered a deep learning architecture. At the same time, if the activation function for a neuron is the step function, it is easy to show that any number of layers can be reduced to two layers (one input and one output layer).
Differing from linear perceptron, MLP can distinguish data that are not linearly separable.
To train the network, the backpropagation algorithm is used, which is a generalization of the least mean squares algorithm in the linear perceptron. 
Backpropagation is used by the gradient descent optimization algorithm to adjust the weight of neurons by calculating the gradient of the loss function~\cite{Mitchell}.

An activation function of a node is a function $f$ defining the output of a node given an input or set of inputs, see Eq.~\eqref{eq:activation}.
In order for ANN to be able to calculate nontrivial functions using a small number of nodes, we need to use nonlinear activation functions. 
\begin{equation}
\label{eq:activation}
y = Activation(\sum(weight \cdot input) + bias).
\end{equation}

In this paper, we consider the logistic (sigmoid) function, tanh function, softmax function, and Rectified Linear Unit function.
The logistic function is a nonlinear function giving smooth and continuously differentiable results~\cite{Haykin}. The range of a sigmoid function is $[0,1]$, which means that all the values going to the next neuron will have the same sign.
\begin{equation}
f(x)=\frac{1}{1+e^{-x}}.
\end{equation}

The tanh function is a scaled version of logistic function where the main difference is that it is symmetric over the origin. The tanh function ranges in $[-1, 1]$.
\begin{equation}
f(x) = tanh(x) = \frac{2}{1+e^{-2x}}-1.
\end{equation}

The softmax function is a type of sigmoid function able to map values into multiple outputs (e.g., classes). The softmax function is ideally used in the output layer of the classifier in order to obtain the probabilities defining a class for each input~\cite{Bishop2006}.

\begin{equation}
f(x)_j = \frac{e^{x_j}}{\sum_{k=1}^{K}e^{x_k}}, for \ j=1,\ldots \ K.
\end{equation}

The Rectified Linear Unit (ReLU) is a nonlinear function that is differing from the previous two activation functions as it does not activate all the neurons at the same time~\cite{Nair}. By activating only a subset of neurons at any time, we make the network sparse and easier to compute. Consequently, such properties make ReLU probably the most widely used activation function in ANNs today.
\begin{equation}
f(x) = max(0, x).
\end{equation}

\subsection{Side-channel Analysis}

Side-channel Analysis (SCA) exploits weaknesses on the implementation level~\cite{dpa_book}. More specifically, all computations running on a certain platform result in unintentional physical leakages. Those leakages are a sort of physical signatures from the reaction time, power consumption, and EM emanations released while the device was manipulating data.
SCA exploits those physical signatures aiming at the key (secret data) recovery.
%Commonly used physical signatures, but not limited to, are power consumption, electromagnetic radiation, timing, cache access patterns etc.
In its basic form, SCA was proposed to perform key recovery attacks on implementation of cryptography~\cite{kocher1996timing,kocher1999differential}.
One advantage of SCA over traditional cryptanalysis is that SCA can apply a divide-and-conquer approach. 
Thus, instead of testing and recovering the full key at once, SCA can be used to recover small parts of the key independently, exponentially reducing the attack complexity. 

However, the scope of SCA is much wider. For example, SCA was recently used to demonstrate IP theft from 3D printers~\cite{faruque2016acoustic}.
Based on the analysis technique different variants of SCA are known.
In the following, we recall a few analysis techniques used later in the paper.
Although the following terms suggest power analysis, these techniques apply to other side-channels as well.

%Side-channel analysis (SCA) is a serious threat, which exploits weakness in physical implementation of cryptographic algorithms rather than the algorithms themselves~\cite{dpa_book}.
%The weakness stems from basic device physics of underlying computing elements i.e., CMOS cells, which makes it hard to eliminate such threats.
%SCA exploits any unintentional leakage observed in physical channels like timing, power dissipation, electromagnetic (EM) radiation, etc.
%For instance, a data transition from $0\rightarrow1$ in a CMOS cell has a different signature than that of $1\rightarrow0$ in power or EM measurement.
%These differences are exploited by an adversary using statistical means, for example, to distinguish correct secret key hypothesis from wrong ones. 
%One advantage of SCA over traditional cryptanalysis is that SCA can apply a divide and conquer approach. 
%Thus, instead of testing and recovering the full key at once, SCA can be used to recover small parts of the key independently, thus exponentially reducing the attack complexity. 

\subsubsection{Simple Power Analysis (SPA)}

Simple power analysis, as the name suggests, is the most basic form of SCA~\cite{kocher1999differential}.
It targets information from a sensitive computation which can be recovered from a single or a few traces.

As a common example, SPA can be used against a straightforward implementation of the RSA algorithm.
Namely, the RSA exponentiation is composed of a sequence of square and multiply operations which depend on secret key bit (multiply follows square only when the secret bit is 1, else only square is executed).
As square and multiply have distinct physical signatures the adversary can directly read out the key bits from e.g., a power trace on a digital oscilloscope.
Similar attacks have been applied to secret-key algorithm like AES~\cite{mangard2002simple} but then targeting key schedule.
In this work, we apply SPA to reverse engineer the architecture of the neural network.

\subsubsection{Differential Power Analysis (DPA)}

DPA is an advanced form of SCA, which applies statistical techniques to recover secret from physical signatures when SPA is not possible.
The attack normally tests for dependencies between actual physical signature (or measurements) and hypothetical physical signature i.e., predictions on intermediate data. The hypothetical signature is based on a leakage model and key hypothesis.
With the divide-and-conquer approach, parts of the secret key (e.g., one byte) can be tested independently, allowing exhaustive search on key hypothesis.
The knowledge of the leakage model comes from the adversary's intuition and expertise.
Some commonly used leakage models for representative devices are the Hamming weight for microcontrollers and the Hamming distance in FPGA, ASIC, and GPU~\cite{bhasin2013cryptography,luo2015side} platforms.
%Apart from the leakage model, the analysis principle remains unchanged.

As the measurements can be noisy, the adversary often needs many measurements, sometimes millions.
Next, statistical tests like correlation~\cite{brier2004correlation} are applied to identify correct key hypothesis from other wrong hypotheses.
%Without loss of generality, we perform our experiments on an 8-bit microcontroller for which high-quality measurements were obtained.
%The choice of device is based on high-quality measurements, however, the methodology can be simply extended to other targets.
As we show later in the paper, DPA is used to recover secret weights from a pre-trained network.

\subsubsection{Horizontal Power Analysis (HPA)}

HPA is another sort of side-channel attack using power as the source of leakage~\cite{clavier2010horizontal}.
While DPA recovers the secret key statistically over multiple measurements, HPA is a single trace attack exploiting several elementary operations in a single computation. The idea behind it is that identical data being manipulated even in different computation steps will have the same power signatures and can be recovered by e.g., pattern recognition techniques.
%For example, in an RSA exponentiation, same operations are computed for every key bit.
HPA can be used against protected implementation, for example with exponent blinding, where an adversary is limited to only one measurement.
%Given the key-size of $n$ bits, there are $n$ square (and multiply) operations in a single measurements.
%To perform HDPA, adversary splits the measurement into $n$ small measurement and apply a standard DPA.
%The higher the value of $n$, the easier the attack becomes.
In this paper, we use HPA to perform input recovery attack for a known network where we prove the technique to be effective for medium to large sized networks.

\section{Side-Channel Based Reverse Engineering of Neural Networks}
\label{sec:experiments}
As already discussed, side-channel leakages have been frequently used for cryptanalysis, in particular for key recovery attacks in cryptography and for the reverse engineering of cryptographic algorithms.
In this work, we demonstrate the first application of SCA for reverse engineering of neural networks, with practical measurements on embedded platforms.

\subsection{Threat Model}

The two main goals of this paper are to recover the neural network architecture and its inputs using only side-channel information.\\
\textbf{Scenario}. We select to work with MLP since 1) it is a commonly used machine learning algorithm in modern applications, see e.g.,~\cite{8053814, 7140247, 7821702, Thomas2015}; 2)  it consists of fully connected layers which are also occurring in other architectures like convolutional neural networks or recurrent neural networks; and 3) the layers are all identical, which makes it more difficult for SCA and could be consequently considered as the worst-case scenario.
We choose our attack to be as generic as possible while discarding common assumptions, which would make the attack easier but also more limited in scope.
For instance, we have no assumption on the type of inputs or its source, as we work with real numbers.
If the inputs are in form of integers (like the MNIST database), the attack becomes easier, since we would not need to recover mantissa bytes and deal with precision.
%In all our experiments, we do not assume any special structure the input data needs to have: consequently, we assume the input data is represented as real values. Note, if we would assume for instance that the input data are integer values, our attack would be much simpler since we would not need to recover mantissa bytes.
We also assume that the implementation of the machine learning algorithm does not include any side-channel countermeasures.
Currently, to the best of our knowledge, no public implementation of ANN deploys side-channel countermeasures.

\textbf{Attacker's capability}. We consider a passive attacker who is only capable of acquiring measurements of the device while operating ``normally'' and not interfering with its operations.
We consider two settings:
\begin{compactenum}
\item Attacker does not know the architecture of the used network but can feed random (or known) inputs to the architecture\\ 

An adequate use case would be when the attacker legally acquires a copy of the network in a black box setting and aims at recovering its internal details, for IP theft. The attacker can query the device with random/chosen inputs and perform side-channel measurements while processing the data. The goal for this setting is to reverse engineer the following information about neural network architecture: number of layers, number of outputs, activation functions, weights in the network.
	\item Attacker knows the architecture but does not know the inputs to it\\ 
	
A suitable use case is where a secret dataset is tested with a public MLP network. The input can correspond to sensitive data such as medical records of patients. The goal for this setting is to obtain the inputs (the data to be classified) to the network and we achieve this with a single measurement only. % In contrast to related works, we do not limit our attack only to the first hidden layer~\cite{DBLP:journals/corr/abs-1803-05847}.
\end{compactenum}

\subsection{Experimental Setup}

\begin{figure}
\centering   
\subfloat[Target 8-bit microcontroller mechanically decapsulated]{\label{f:sa}\includegraphics[width=0.5\linewidth]{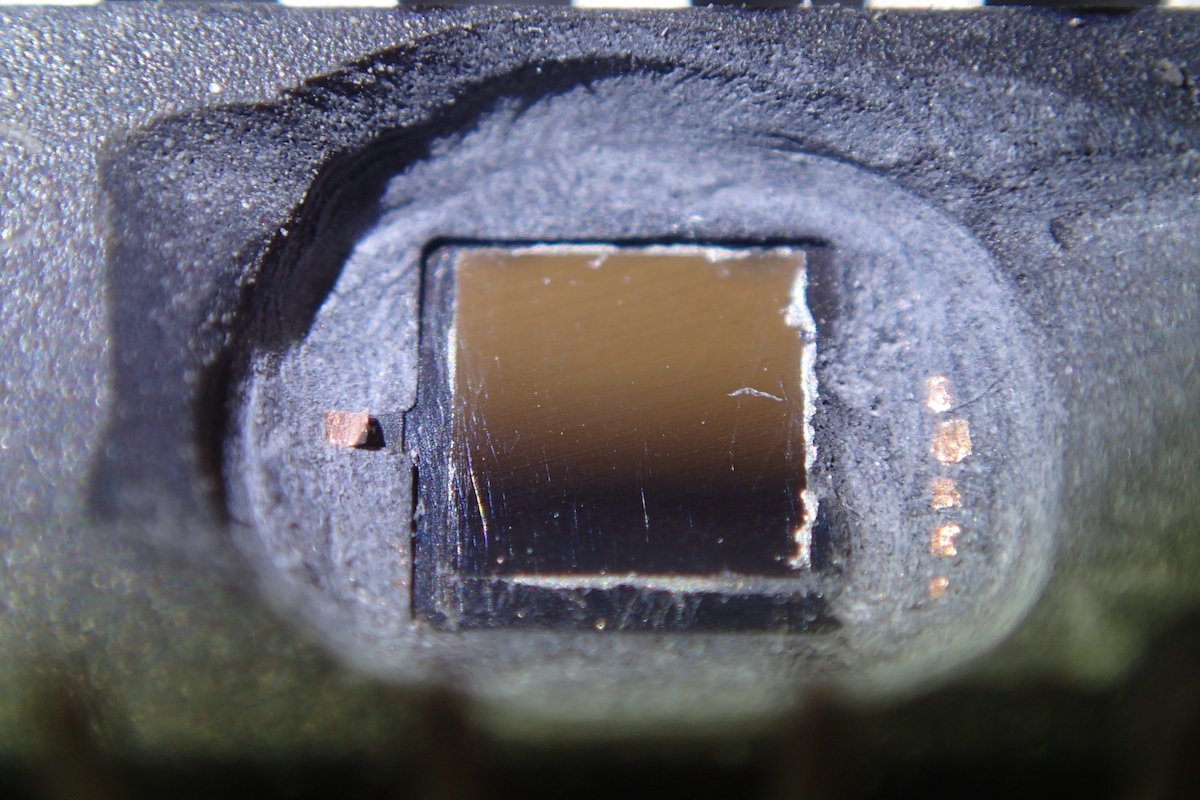}}
\subfloat[Langer RF-U 5-2 Near-field Electromagnetic passive Probe]{\label{f:sb}\includegraphics[width=0.444\linewidth]{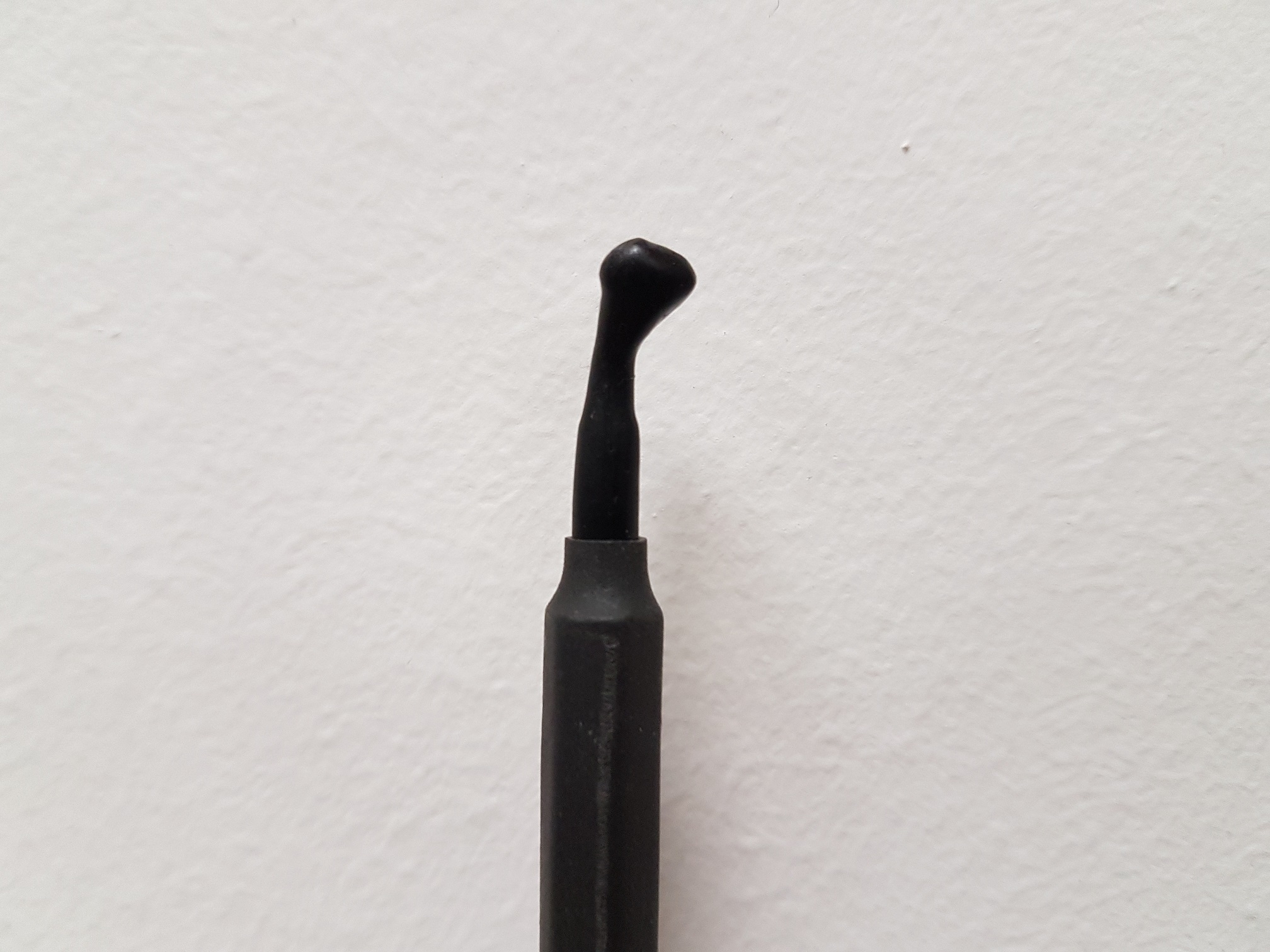}}\\
\subfloat[The complete measurement setup]{\label{f:sc}\includegraphics[width=0.9\linewidth]{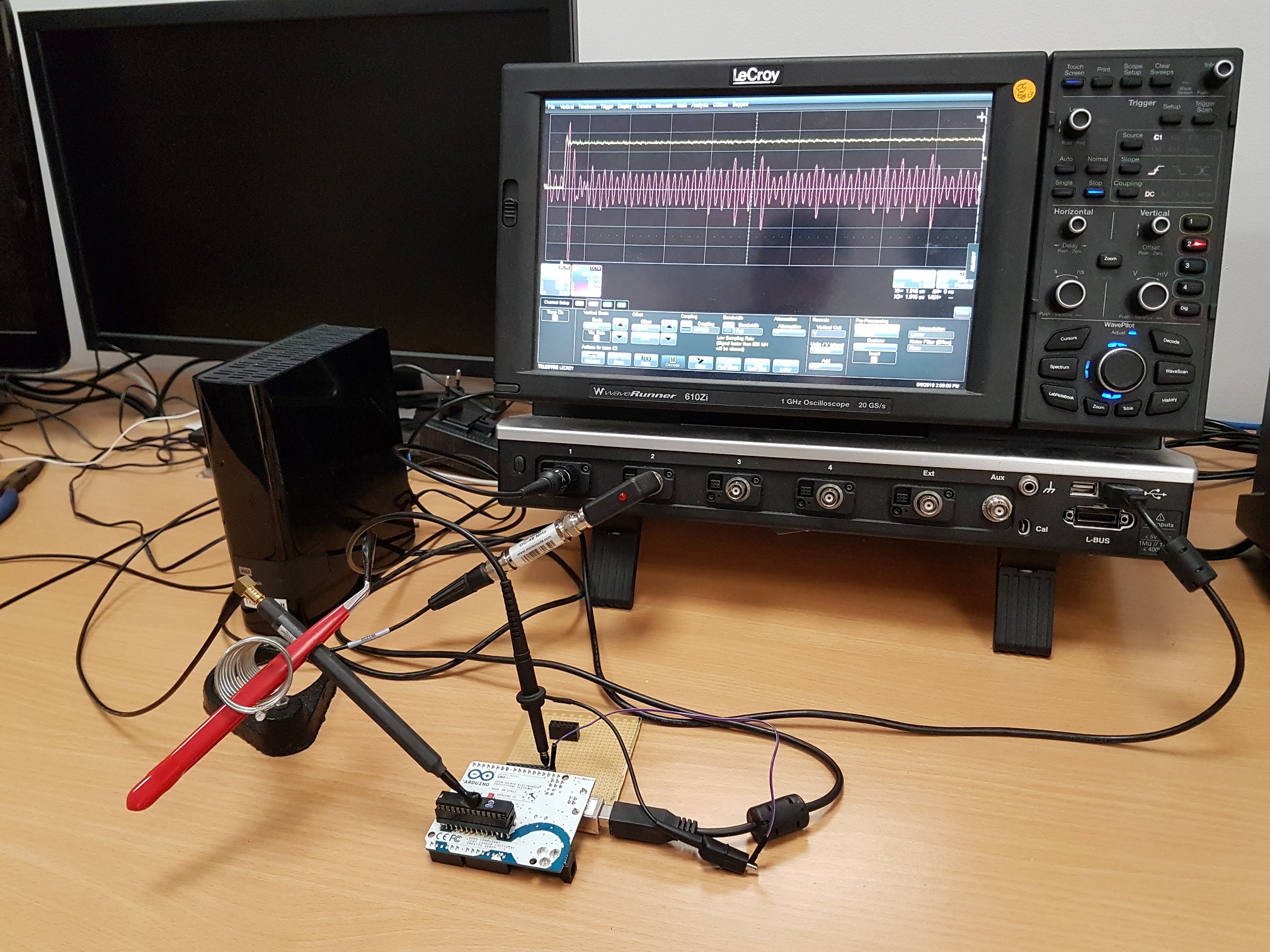}}
	\caption{Experimental Setup}
\label{f:s}
\end{figure}

Here we describe the attack methodology, which is first validated on Atmel ATmega328P.
Later, we also demonstrate the proposed methodology on ARM Cortex-M3.
The side-channel measurements are collected  during the execution of the classification and they are captured using the Lecroy WaveRunner 610zi oscilloscope.
The oscilloscope measurements are synchronized with the operations by common hand shaking signals like start and stop of computation.
To further improve the quality of measurements, we opened the chip package mechanically (see Figure~\ref{f:sa}).
An RF-U 5-2 near-field electromagnetic (EM) probe from Langer is used to collect the measurements (see Figure~\ref{f:sb}).
%The ideal positioning of the probe is determined by benchmarking with the AES measurements on the same platform.
Note that EM measurements also allow to observe the timing of all the operations and thus the setup allows for timing side-channels based analysis as well.
The setup is depicted in Figure~\ref{f:sc}.

Our choice of the target platform is motivated by:
\begin{itemize}
	\item Atmel ATmega328P: This processor allows for high quality measurements. We are able to achieve a high signal-to-noise ratio (SNR) measurements, allowing us to focus on developing the methodology.
	\item ARM Cortex-M3: A modern 32-bit microcontroller architecture with multiple stages of pipeline, on chip co-processors, low SNR measurements, and wide application. We show that the developed methodology is indeed versatile across targets with a  relevant update of measurement capability.
\end{itemize}
For different platforms, the leakage model could change, but this would not limit our approach and methodology.
In fact, those leakage models are well known for other common platforms like FPGA~\cite{bhasin2013cryptography} and GPU~\cite{luo2015side}.
Moreover, as for ARM Cortex-M3, low SNR of the measurement might force the adversary to increase the number of measurements and apply signal pre-processing techniques, but the principles of the analysis remain valid.

%In this section, we elaborate on how side-channel leakage could be exploited in order to perform reverse engineering on the network layout.
%In the previous section, it has been described on different information could be recovered by observing different physical leakage of the device.
%As an example, we conducted the experiments on a simple target, namely Atmel ATmega328P, mounted on Arduino UNO development board.
%We use simple target, since it is easier to highlight the information leakage due to higher signal-to-noise ratio (SNR), and hence, simpler analysis could be performed. Nevertheless, the same principle can be used to target more complex architecture, albeit requiring more measurements and pre-processing.
As already stated above, the exploited leakage model of the target device is the Hamming weight (HW) model.
A microcontroller loads sensitive data to a data bus to perform indicated instructions.
This data bus is pre-charged to all '0's' before every instruction.
Note that data bus being pre-charged is a natural behavior of the microcontroller and not a vulnerability introduced by the attacker.
Thus, the new power consumption (or EM radiation) is modeled as the number of bits equal to '1' in the loaded data.
In other words, the power consumption of loading data $x$ is:
\begin{align}
	HW(x) = \sum_{i=1}^{n} x_i \enspace,
\end{align}
where $x_i$ represents the $i^{th}$ bit of $x$. 
In our case, it is the secret pre-trained weight which is regularly loaded from memory for processing and results in the HW leakage.
To conduct the side-channel analysis, we perform the divide-and-conquer approach, where we target each operation separately. 
The full recovery process is described in Section~\ref{sec:recover}.

%The network is downloaded to the microcontoller board running with a clock frequency of 16 MHz. For measuring the EM traces, Lecroy WaveRunner 610zi oscilloscope is used at a sampling rate of 1 GSamples/s. The EM probe is positioned over the chip with the package mechanically opened, and the probe position is heuristically determined first by trial and error, and later verified by some test attack code for measurement correctness. Even though we focused on measuring the EM leakage, by performing the SPA on the traces collected, the timing behavior can also be observed.
 
Several pre-trained networks are implemented on the board.
The training phase is conducted offline, and the trained network is then implemented in C language and compiled on the microcontroller.
In our experiments, we consider multilayer perceptron architectures consisting of a different number of layers and nodes in those layers. Note that, with our approach, there is no limit in the number of layers or nodes we can attack, as the attack scales linearly with the size of the network. 
The methodology is developed to demonstrate that the key parameters of the network, namely the weights and activation functions can be reverse engineered.
Further experiments are conducted on deep neural networks with three hidden layers. We emphasize that the method we use can be applied to larger networks as well. 
%We further showed that the approach can also be extended to multi-layer networks, making it possible to target deep neural networks.

\subsection{Reverse Engineering the Activation Function}

We remind the reader that nonlinear activation functions are necessary in order to represent nonlinear functions with a small number of nodes in a network. As such, they are elements used in virtually any neural network architecture today~\cite{Krizhevsky:2012:ICD:2999134.2999257, DBLP:journals/corr/HeZRS15}. If the attacker is able to deduce the information on the type of used activation functions, he/she can use that knowledge together with information about input values to deduce the behavior of the whole network.

\begin{figure}
\centering   
\includegraphics[width=0.8\linewidth]{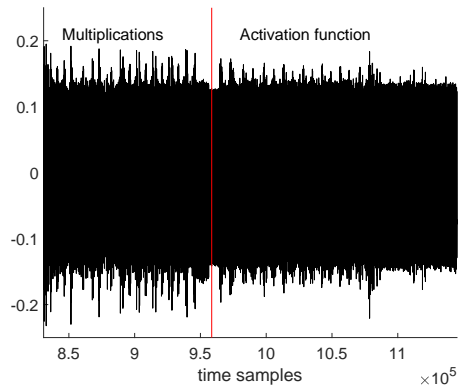}
	\caption{Observing pattern and timing of multiplication and activation function}
\label{onel}
\end{figure}

We analyze the side-channel leakage from different activation functions.
We consider the most commonly used activation functions, namely ReLU, sigmoid, tanh, and softmax~\cite{Haykin,Nair}.
The timing behavior can be observed directly on the EM trace.
For instance, as shown later in Figure~\ref{f:3a}, a multiplication is followed by activation with individual signatures.
For a similar architecture, we test different variants with each activation function.
We collect EM traces and measure the timing of the activation function computation from the measurements.
% for those operations during the profiling step.\todo{what profiling step??}
%\todo{profiling step is not defined}
The measurements are taken when the network is processing random inputs in the range, i.e., $x \in \{-2,2\}$.
A total of $2\,000$ EM measurements are captured for each activation function.
%The range might be small, but it provides enough information to characterize the activation functions.
As shown in Figure~\ref{onel}, the timing behavior of the four tested activation functions have distinct signatures allowing easy characterization.
%Our preliminary observation is that we could not distinguish the input of the functions directly from the leakage, however, based on the timing delay, it is possible to characterize different functions.

\begin{figure}
\centering   
\subfloat[ReLU]{\label{f:0a}\includegraphics[width=0.5\linewidth]{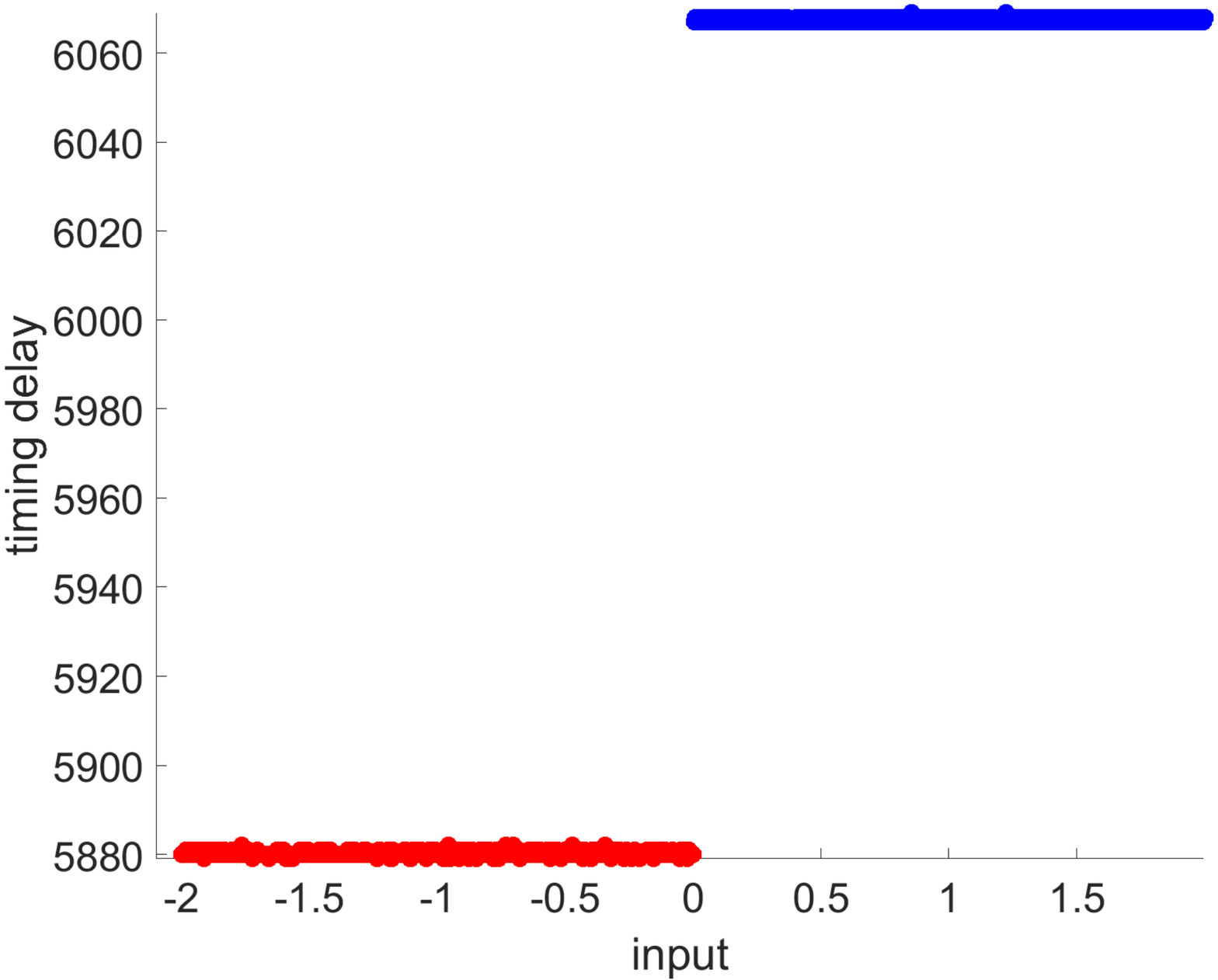}}
\subfloat[Sigmoid]{\label{f:0b}\includegraphics[width=0.5\linewidth]{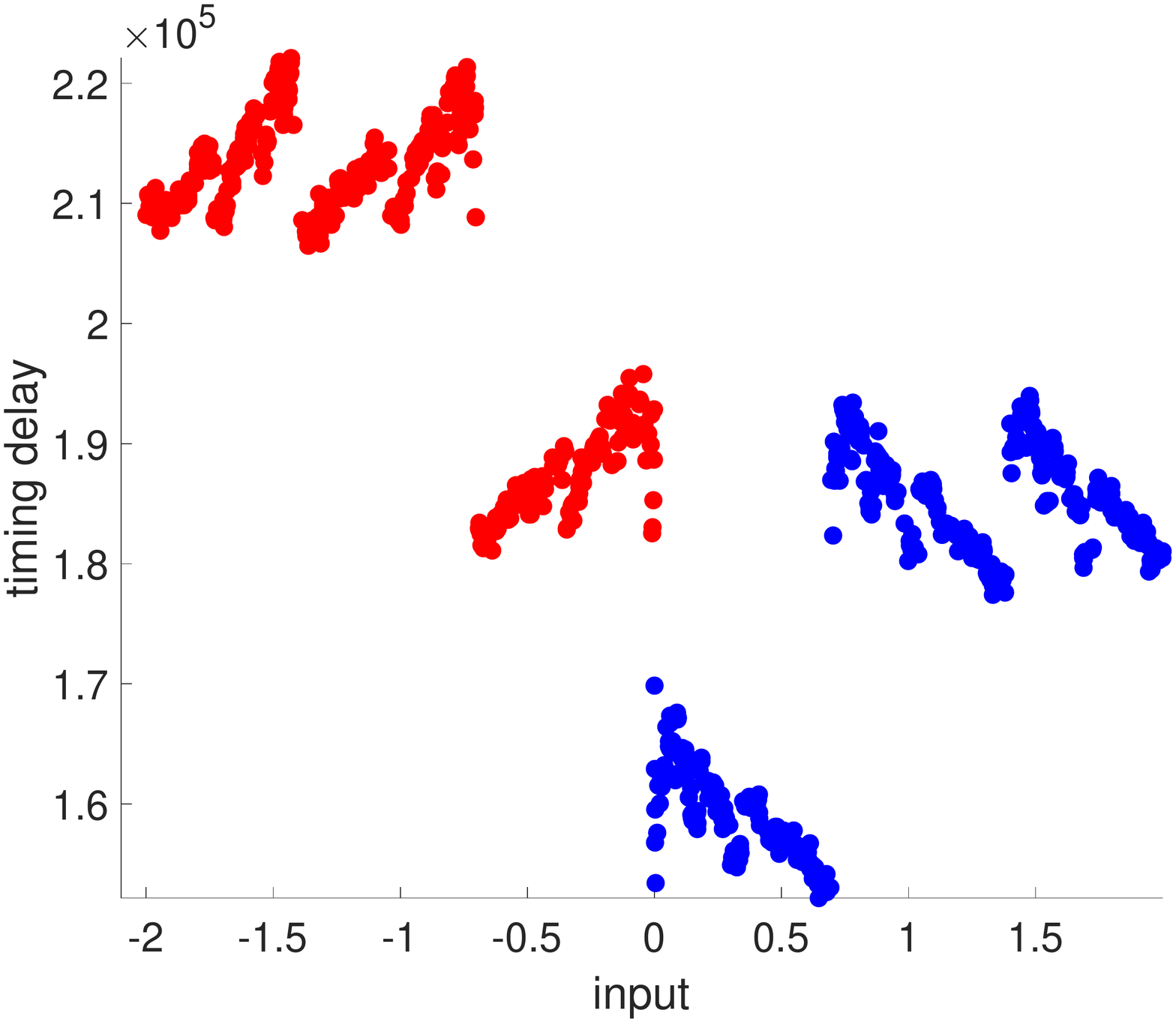}}\\
\subfloat[Tanh]{\label{f:0c}\includegraphics[width=0.5\linewidth]{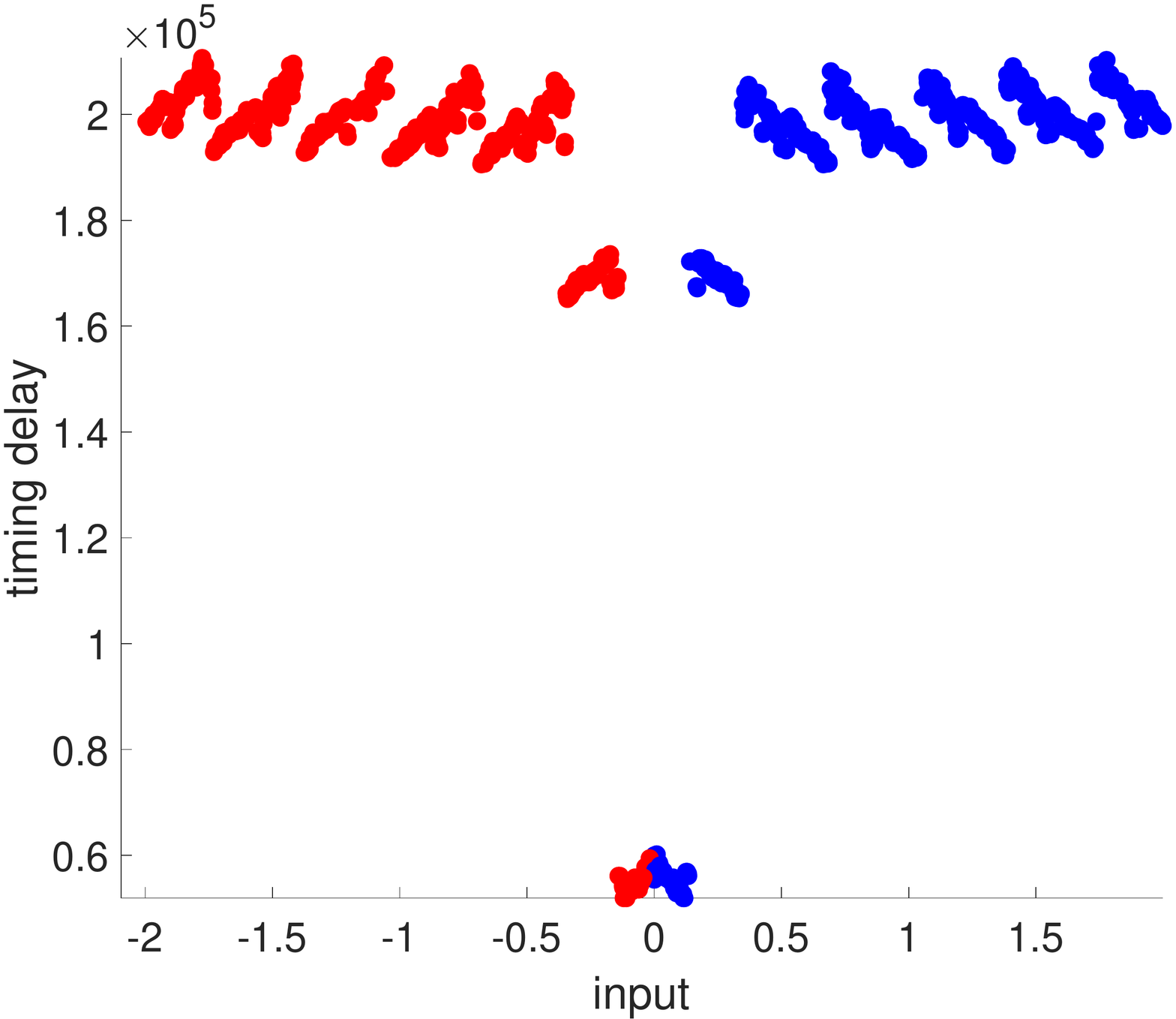}}
\subfloat[Softmax]{\label{f:0d}\includegraphics[width=0.5\linewidth]{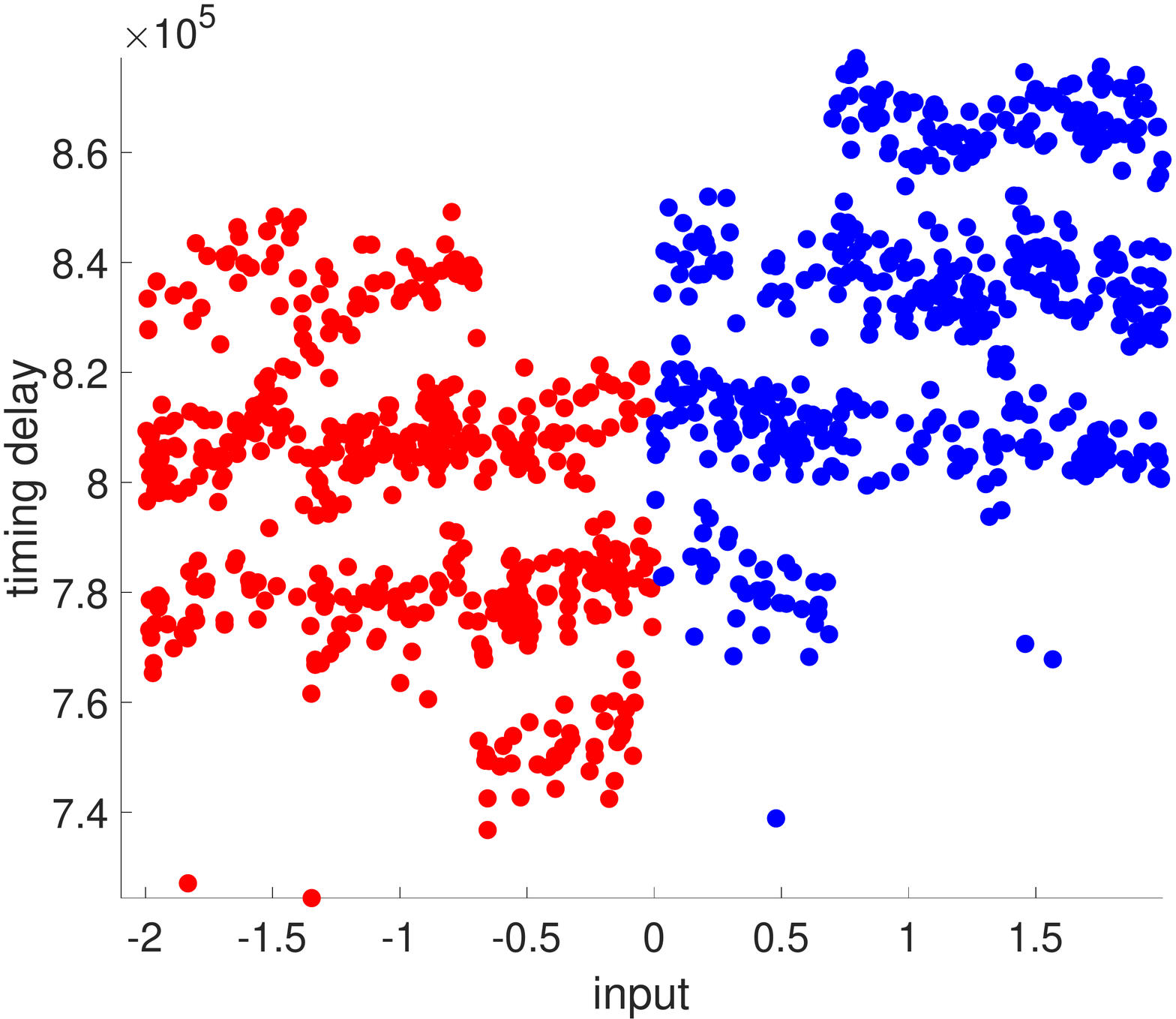}}\\
\caption{Timing behavior for different activation functions}
\label{f:0}
\end{figure}

%In Figure~\ref{f:0}, it can be observed that for different activation functions,
Different inputs result in different processing times. % behavior is quite different. 
%To understand the results better, it will require different timing delays for different input values.
%Although, the input values might not be directly recovered, the timing information can be used to characterize each function.
Moreover, the timing behavior for the same inputs largely varies depending on the activation function.
For example, we can observe that ReLU will require the shortest amount of time, due to its simplicity (see Figure~\ref{f:0a}).
On the other hand, tanh and sigmoid might have similar timing delays, but with different pattern considering the input (see Figure~\ref{f:0b} and Figure~\ref{f:0b}), where tanh is more symmetrical in pattern compared to sigmoid, for both positive and negative inputs. We can observe that softmax function will require most of the processing time, since it requires the exponentiation operation which also depends on the number of neurons in the output layer.
As neural network algorithms are often optimized for performance, the presence of such timing side-channels is often ignored.
A function such as tanh or sigmoid requires computation of $e^x$ and division and it is known that such functions are difficult to implement in constant time.
In addition, constant time implementations might lead to a substantial performance degradation.
%, in naive implementation, it might take much longer time and hence, any bias in timing delay might be more observable.
Other activation functions can be characterized similarly.
Finally, Table~\ref{tab:time} presents the minimum, maximum, and mean computation time for each activation function over captured $2\,000$ measurements.
While ReLU is fastest, the timing difference of each function stands out individually, thus allowing a straightforward recovery.

\begin{table}
	\caption{Minimum, Maximum, and Mean computation time (in $ns$) for different activation functions}
	\label{tab:time}
	\centering
	\begin{tabular}{c|ccc}
		\hline
		Activation Function & Minimum  & Maximum  &   Mean   \\ \hline
		       ReLU         &  5\,879  &  6\,069  &  5\,975  \\
		      Sigmoid       & 152\,155 & 222\,102 & 189\,144 \\
		       Tanh         & 51\,909  & 210\,663 & 184\,864 \\
		      Softmax       & 724\,366 & 877\,194 & 813\,712 \\ \hline
	\end{tabular}
\end{table}

\subsection{Reverse Engineering of the Multiplication Operation}\label{sec:weight}

\begin{figure*}
\centering   
\subfloat[First byte mantissa for weight = 2.43]{\label{f:2a}\includegraphics[width=0.33\textwidth]{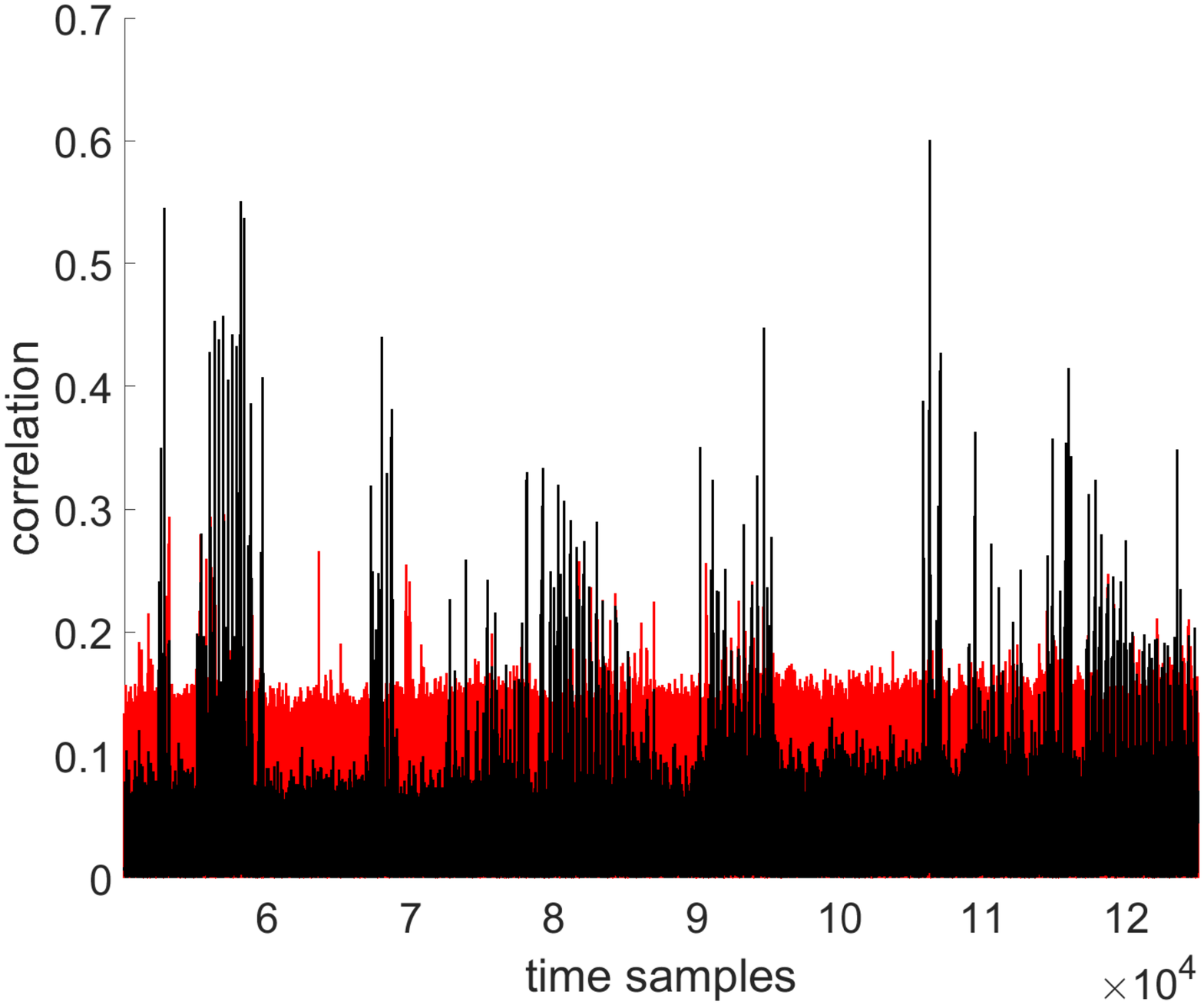}}
\subfloat[Second byte mantissa for weight = 2.43]{\label{f:2b}\includegraphics[width=0.33\textwidth]{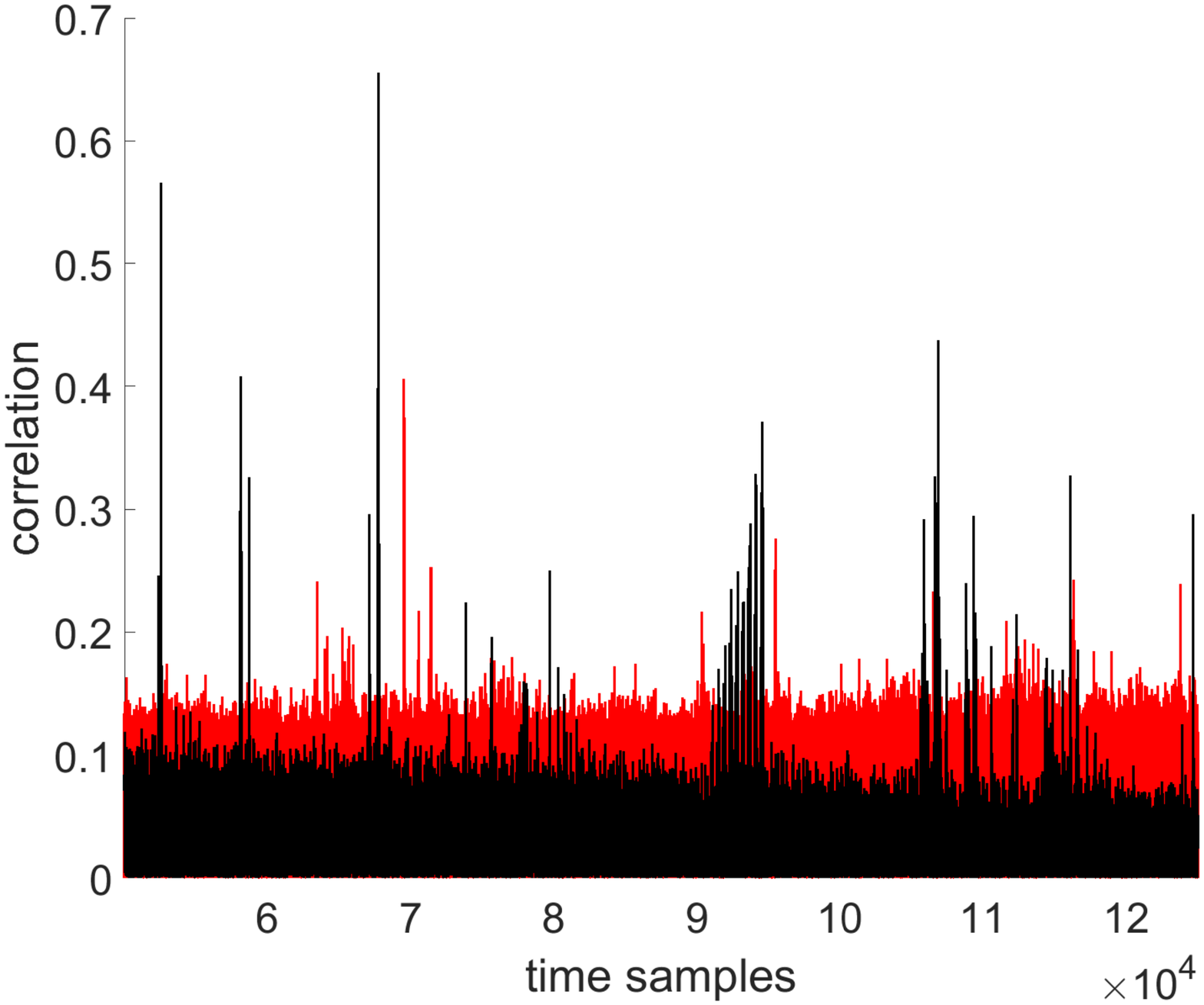}}
\subfloat[Third byte mantissa for weight = 2.43]{\label{f:2c}\includegraphics[width=0.33\textwidth]{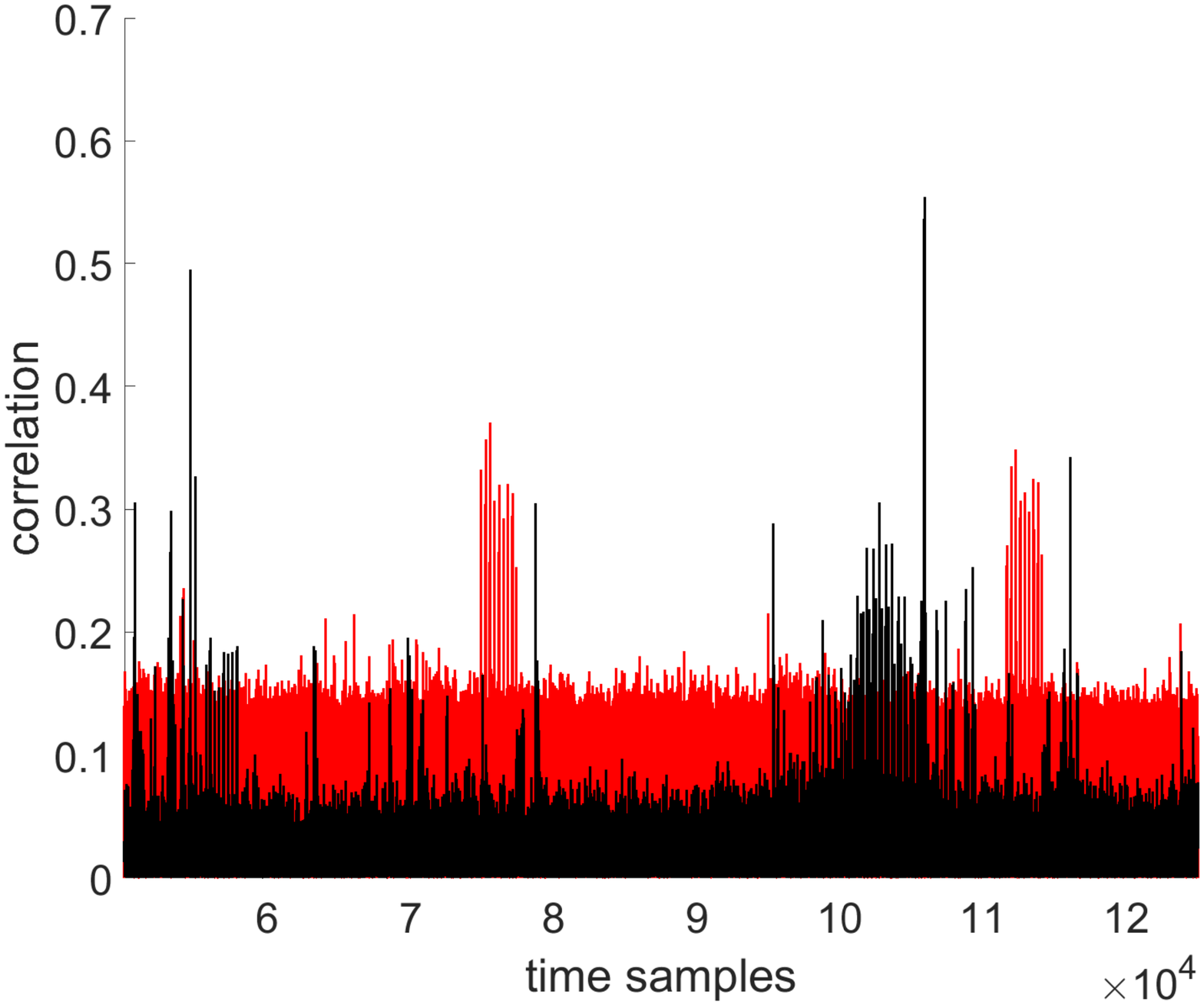}}\\
\caption{Correlation of different weights candidate on multiplication operation}
\label{f:2}
\end{figure*}

\begin{figure}
\centering   
\subfloat[weight = 1.635]{\label{f:2e}\includegraphics[width=0.8\linewidth]{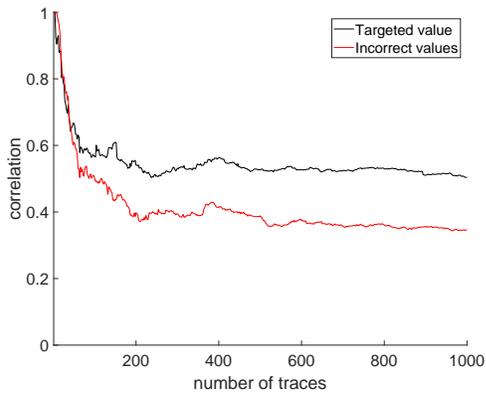}}\\
\subfloat[weight = 0.890]{\label{f:2f}\includegraphics[width=0.8\linewidth]{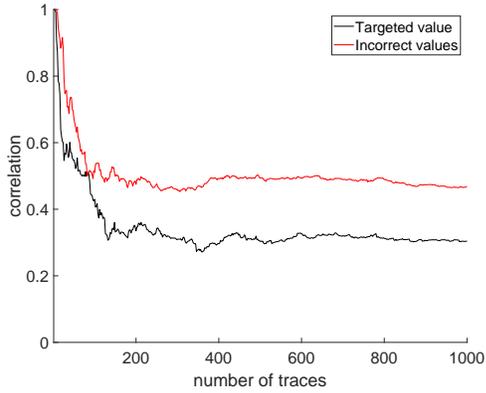}}\\
\caption{Correlation comparison between correct and incorrect mantissa of the weights}
\label{f:2z}
\end{figure}

A well-trained network can be of a significant value.
What distinguishes a good versus poorly trained network for a given architecture are the weights.
With fine-tuned weights, we can improve the accuracy of the network, which has both commercial and academic interest.
In the following, we demonstrate a way to recover those weights by using SCA.

For the recovery of the weights, we use the Correlation Power Analysis (CPA) i.e., a variant of DPA using the Pearson's correlation as a statistical test.
CPA targets the multiplication $m = x \cdot w$ of a known input $x$ with a secret weight $w$.
Using the HW model, the adversary correlates the activity of the predicted output $m$ for all hypothesis of the weight.
Thus, the attack computes $\rho(t,w)$, for all hypothesis of the weight $w$, where $\rho$ is the Pearson correlation coefficient and $t$ is the side-channel measurement.
The correct value of the weight $w$ will result in a higher correlation standing in this way out from all other wrong hypotheses $w^*$, given enough measurements.
Although the attack concept remains the same as in the case of an attack on cryptographic ciphers, the actual attack used here is quite different.
While cryptographic operations are always performed on fixed length integers, in ANN we are dealing with real numbers. % which might make the analysis more complicated.

%In a classical CPA on 
%The concept is similar, in CPA, the attack is aimed at the output of $f(x\oplus w)$, where $x$ is known input value, $w$ is unknown secret value, and $f$ is non-linear function. In the case of weight recovery, similar principle applies, where we target $x \cdot w$ \textcolor{red}{[any ref on attack on multiplication?]} and the unknown secret value $w$ is the weight. The difference is that for side-channel attack, the target is cryptographic operations, which normally operate on integer values of fixed bit length. On the other hand, in neural network, all the operations are mostly dealing with real numbers which might make the analysis more complicated (Although fixed point approach can be used when performing the computation, but in this case, it reduces to basic side-channel attack with unbounded bit length \textcolor{red}{(?)}).
%

We start by analyzing the way the compiler is handling floating-point operations for our target.
The generated assembly is shown in Table~\ref{t:asm}, which confirms the usage of IEEE 754 compatible representation as stated above.
The knowledge of the representation allows one to better estimate the leakage behavior.
Since the target device is an 8-bit microcontroller,  the representation follows 32-bit pattern $(b_{31}...b_{0})$, which is stored in 4 registers.
The 32-bit consist of: 1 sign bit $(b_{31})$, 8 biased exponent bits $(b_{30}...b_{23})$ and 23 mantissa (fractional) bits $(b_{22}...b_{0})$. It can be formulated as: 
%
%For our target device, we then investigate how it processes floating point operation, by checking the generated assembly code by the compiler (see Table~\ref{t:asm}), in order to model the leakage behavior. 
%Based on the generated assembly, we assume that the floating point will be represented as IEEE 754 format \textcolor{red}{[put reference?]}.
%The IEEE Standard for Floating-Point Arithmetic (IEEE 754) is a standard for floating point computation. Normally, in most microcontroller, when dealing with real number on high level language, the value has to be stored in binary representation, which in then standardized in IEEE 754.  In our 8-bit microcontroller case,  the representation will most likely be a 32-bit representation $(b_{31}...b_{0})$, which is stored in 4 registers. The 32-bit consists of: 1 sign bit $(b_{31})$, 8 biased exponent bits $(b_{30}...b_{23})$ and 23 mantissa (fractional) bits $(b_{22}...b_{0})$. It can be formulated as: 
\begin{equation*}
(-1)^{b_{31}}\times 2^{(b_{30}...b_{23})_2 - 127}\times (1.b_{22}...b_{0})_2.
\end{equation*}
For example, the value $2.43$ can be expressed as $(-1)^{0}\times 2^{(1000000)_2 - 127}\times (1.00110111000010100011111)_2$.
The measurement $t$ is considered when the computed result $m$ is stored back to the memory, leaking in the HW model i.e., $HW(m)$.
Since 32-bit $m$ is split into individual 8-bits, each byte of $m$ is recovered individually.
%We thus target the leakage during the storing of the result back to the memory, and for the leakage model, we can use the Hamming weight (HW) leakage model, commonly used for 8-bit microcontroller.
Hence, by recovering this representation, it is enough to recover the estimation of the real number value.

\begin{table}[h]
\centering
\caption{Code snippet of the returned assembly for multiplication: $x = x\cdot w(=2.36$ or 0x3D0A1740 in IEEE 754 representation). The multiplication itself is not shown here, but from the registers assignment, our leakage model assumption holds.}
\label{t:asm}
	%\begin{scriptsize}
\ttfamily
\begin{tabular}{p{1cm} p{3cm}  p{3cm}}
	\textbf{\#} & \textbf{Instruction} & \textbf{Comment}
	\\
	\hline
 11a &	ldd	r22, Y+1	& 0x01\\
 11c  &   ldd	r23, Y+2	&0x02\\
 11e&	ldd	r24, Y+3&	0x03\\
 120&	ldd	r25, Y+4&	0x04\\
 122&	ldi	r18, 0x3D&	61\\
 124&	ldi	r19, 0x0A&	10\\
 126&	ldi	r20, 0x17&	23\\
 128	&ldi	r21, 0x40& 64\\
 12a	&call	0xa0a	& multiplication\\
 12e&	std	Y+1, r22&	0x01\\
 130&	std	Y+2, r23&	0x02\\
 132&	std	Y+3, r24&	 0x03\\
 134&	std	Y+4, r25	&0x04\\
\hline
\end{tabular}
%	\end{scriptsize}
\end{table}

To implement the attack two different approaches can be considered.
The first approach is to build the hypothesis on the weight directly.
For this experiment, we target the result of the multiplication $m$ of known input values $x$ and unknown weight $w$.
For every input, we assume different possibilities for weight values. We then perform the multiplication and estimate the IEEE 754 binary representation of the output.
To deal with the growing number of possible candidates for the unknown weight $w$, we assume that the weight will be bounded in a range $[-N,N]$, where $N$ is a parameter chosen by the adversary, and the size of possible candidates is denoted as $s = 2N/p$, where $p$ is the precision when dealing with floating-point numbers.

Then, we perform the recovery of the 23-bit mantissa of the weight. The sign and exponent could be recovered separately.
Thus, we are observing the leakage of 3 registers, and based on the best CPA results for each register, we can reconstruct the mantissa.
Note that the recovered mantissa does not directly relate to the weight, but with the recovery of the sign and exponent, we could obtain the unique weight value.
%Since the attack can be performed in divide-and-conquer approach, we implement the multiplication operation in C and we then try to perform the operation.
The traces are measured when the microcontroller performs secret weight multiplication with uniformly random values between -1 and 1 ($x \in \{-1,1\}$) to emulate normalized input values.
We set $N = 5$ and to reduce the number of possible candidates, we assume that each floating-point value will have a precision of 2 decimal points, $p = 0.01$. Since we are dealing with mantissa only, we can then only check the weight candidates in the range $[0,N]$, thus reducing the number of possible candidates.

In Figure~\ref{f:2}, we show the result of the correlation for each byte with the measured traces.
The horizontal axis shows time of execution and vertical axis correlation.
%The correlation of correct value (shown in black), stands out of all wrong hypotheses (shown in red).
The experiments were conducted on 1\,000 traces for each case. In the figure, the black plot denotes the correlation of the ``correct'' mantissa weight ($|m(\hat{w}) - m(w)| <0.01$), whereas the red plots are from all other weight candidates in the range described earlier.  Since we are only attacking mantissa in this phase, several weight candidates might have similar correlation peaks.
After the recovery of the mantissa, the sign bit and exponent can be recovered similarly, which narrows down the list candidate to a unique weight.
Another observation is that the correlation value is not very high and scattered across different clock cycles.
This is due to the reason that the measurements are noisy and since the operation is not constant-time, the interesting time samples are distributed across multiple clock cycles. Nevertheless, it is shown that the side-channel leakage can be exploited to recover the weight up to certain precision.
Multivariate side channel analysis~\cite{prouff2013masking} can be considered if distributed samples hinder recovery.

We emphasize that attacking real numbers as in the case of weights of ANN can be simpler than attacking cryptographic implementations.
This is because cryptography works on fixed length integers and exact values must be recovered.
When attacking real numbers, small precision errors due to rounding off the intermediate values still result in useful information. 

To deal with more precise values, we can target the mantissa multiplication operation directly. In this case, the search space can either be $[0,2^{23}-1]$ to cover all possible values for the mantissa (hence, more computational resources will be required) or we can focus only on the most significant bits of the mantissa (lesser candidates but also with lesser precision). Since the 7 most significant bits of the mantissa are processed in the same register, we can aim to target only those bits, assigning the rest to 0. Thus, our search space is now $[0, 2^7-1]$.
The mantissa multiplication can be performed as $1.mantissa_x \times 1.mantissa_w$, then taking the 23 most significant bits after the leading 1, and normalization (updating the exponent if the result overflows) if necessary. 

In Figure~\ref{f:2z}, we show the result of the correlation between the HW of the first 7-bit mantissa of the weight with the traces. Except for Figure~\ref{f:2f}, the other results show that the correct mantissa can be recovered.
The most interesting result is shown in Figure~\ref{f:2f}, which at the first glance looks like a failure of the attack.
Here, the target value of the mantissa is \textbf{1100011}110...10, while the attack recovers \textbf{1100100}000..00.
Considering the sign and exponents, the attack recovers \emph{0.890625} instead of \emph{0.89}, i.e., a precision error at $4^{th}$ place after decimal point.
%Unlike SCA on cryptography, such precision errors in recovery still leads to a useful value. 
%For the case of Figure~\ref{f:2f}, we checked that the targeted mantissa is \textbf{1100011}110...10, and the recovered value is \textbf{1100100}000..00. If the sign and exponent bits are recovered correctly, we get the value 0.890625, which is not a far from the actual value 0.89. This might due to the reason that we only restricted the precision to 7 bit.
Thus, in both cases, we have shown that we can recover the weights from the SCA leakage.

\begin{figure}
\centering   
\subfloat[First byte recovery (sign and 7-bit exponent)]{\label{f:2e2}\includegraphics[width=0.8\linewidth]{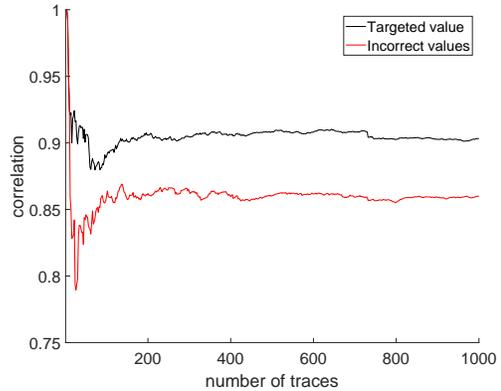}}\\
\subfloat[Second byte recovery (lsb exponent and mantissa)]{\label{f:2f2}\includegraphics[width=0.8\linewidth]{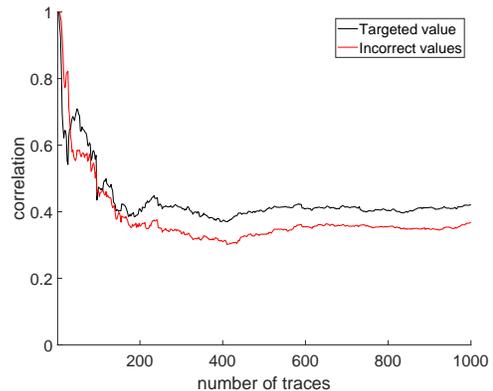}}\\
\caption{Recovery of the weight}
\label{f:2z2}
\end{figure}

Lastly, in Figure~\ref{f:2z2}, we show the composite recovery of 2 bytes of the weight representation i.e., a low precision setting where we recover sign, exponent and most significant part of mantissa. Again, the targeted (correct) weight can be easily distinguished from the other candidates.
Hence, once all the necessary information has been recovered, the weight can be reconstructed accordingly.

\subsection{Reverse Engineering the Number of Neurons and Layers}

After the recovery of the weights and the activation functions, in this step, we use SCA to determine the structure of the network.
Mainly, we are interested to see if we can recover the number of hidden layers and the number of neurons for each layer.
To perform the reverse engineering of the network structure, we first use SPA.
SPA is the simplest form of SCA which allows information recovery in a single (or a few) traces with methods as simple as visual inspection.
The analysis is performed on three networks with different layouts.

\begin{figure*}
\centering   
\subfloat[One hidden layer with 6 neurons]{\label{f:3a}\includegraphics[width=0.33\linewidth]{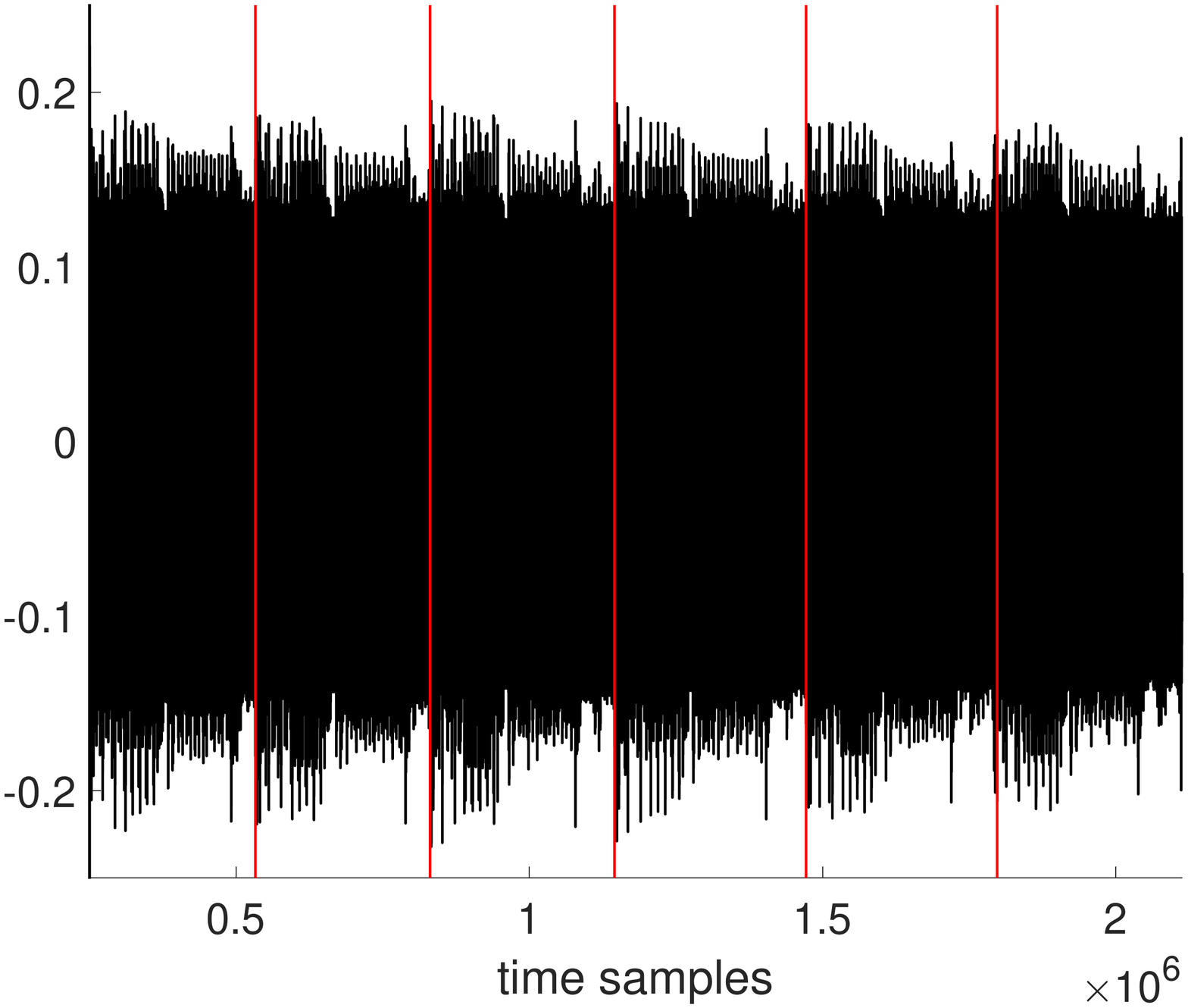}}
\subfloat[2 hidden layers (6 and 5 neurons each)]{\label{f:3b}\includegraphics[width=0.33\linewidth]{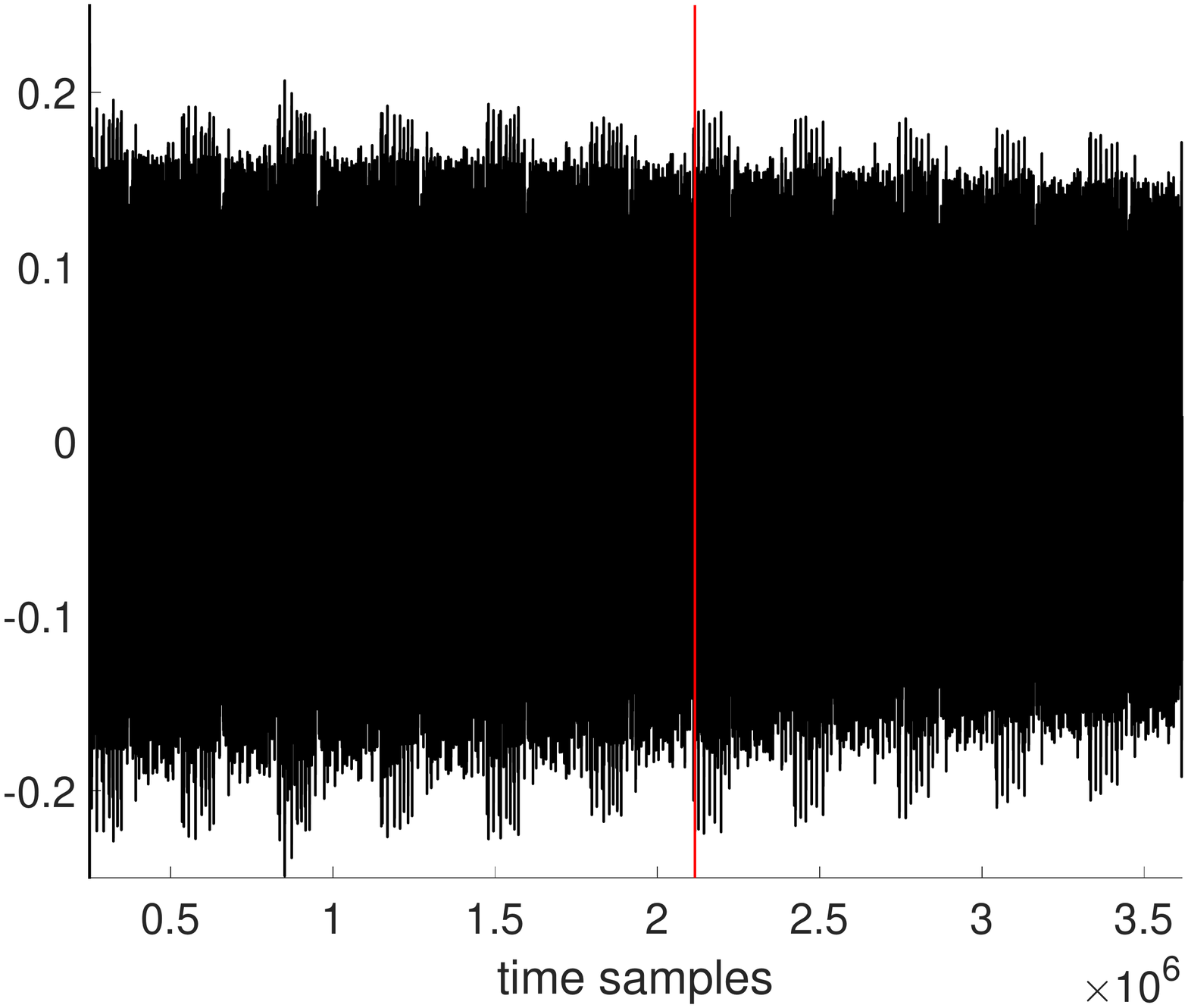}}
\subfloat[3 hidden layers (6,5,5 neurons each)]{\label{f:3c}\includegraphics[width=0.33\linewidth]{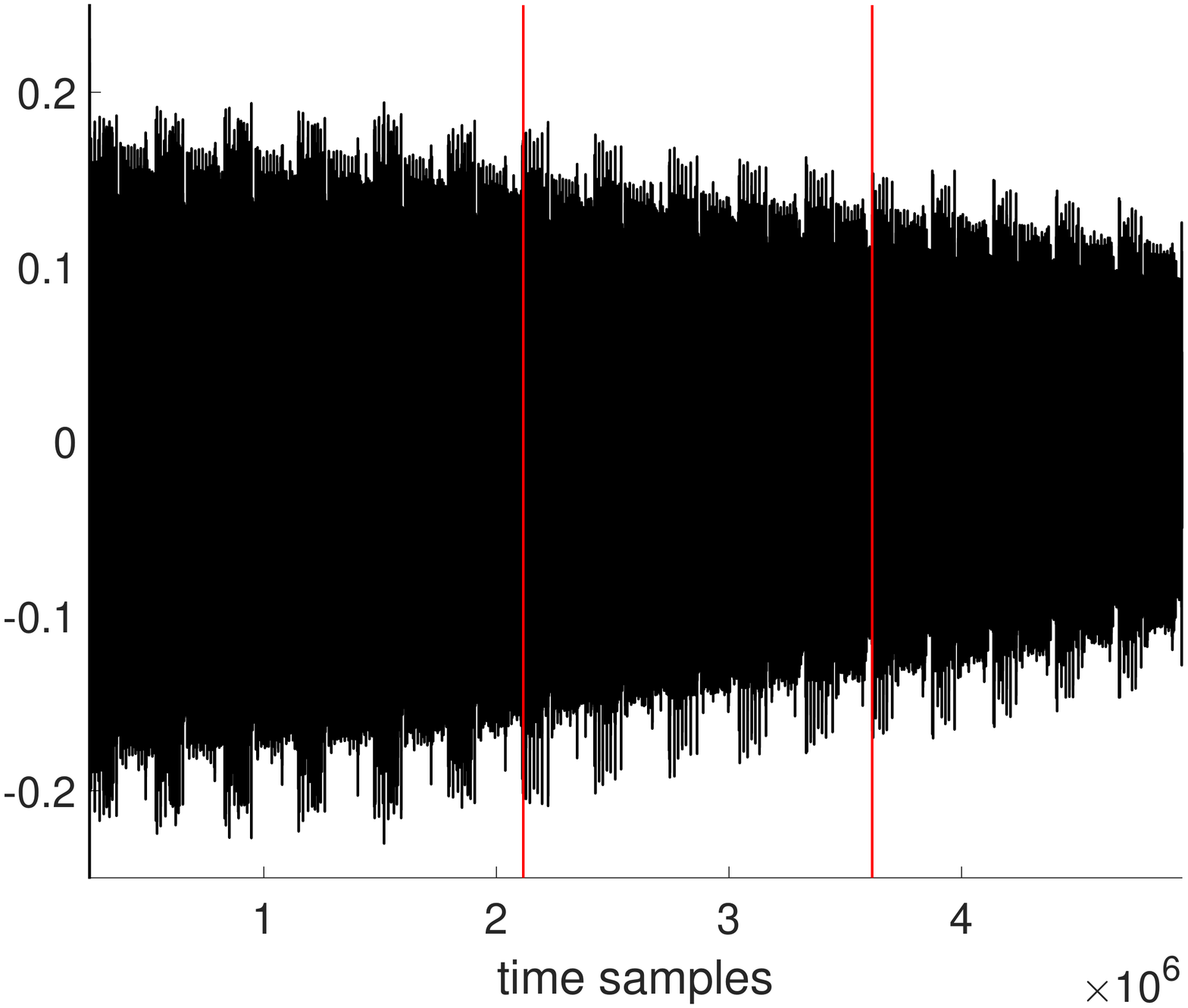}}
\caption{SPA on hidden layers}
\label{f:3}
\end{figure*}

The first analyzed network is an MLP with one hidden layer with 6 neurons. 
The EM trace corresponding to the processing of a randomly chosen input is shown in Figure~\ref{f:3a}.
By looking at the EM trace, the number of neurons can be easily counted.
The observability arises from the fact that multiplication operation and the activation function (in this case, it is the Sigmoid function) have completely different leakage signatures.
Similarly, the structures of deeper networks are also shown in Figure~\ref{f:3b} and Figure~\ref{f:3c}.
The recovery of output layer then provides information on the number of output classes.
However, distinguishing different layers might be difficult, since the leakage pattern is only dependent on multiplication and activation function, which are usually present in most of the layers.
We observe minor features allowing identification of layer boundaries but only with low confidence.
Hence, we develop a different approach based on CPA to identify layer boundaries. % \textcolor{blue}{I dont think the last comment is true. its just not true for our experiments. This should be handled carefully}.

%We apply advanced DPA technique to distinguish the different layers.
The experiments follow similar methodology as in the previous experiments.
To determine if the targeted neuron is in the same layer as previously attacked neurons, or in the next layer, we perform a weight recovery using two sets of data.

Let us assume that we are targeting the first hidden layer (the same approach can be done on different layers as well). 
Assume that the input is a vector of length $N_0$, so the input $x$ can be represented $x = \{x_1, ..., x_{N_0}\}$.
For the targeted neuron $y_n$ in the hidden layer, perform the weight recovery on 2 different hypotheses.
For the first hypothesis, assume that the $y_n$ is in the first hidden layer. Perform weight recovery individually using $x_i$, for $1 \leq i \leq N_0$.
For the second hypothesis, assume that $y_n$ is in the next hidden layer (the second hidden layer). Perform weight recovery individually using $y_i$, for $1 \leq i \leq (n-i)$. For each hypothesis, record the maximum (absolute) correlation value, and compare both. Since the correlation depends on both inputs to the multiplication operation, the incorrect hypothesis will result in a lower correlation value. Thus, this can be used to identify layer boundaries. %at which neuron, it starts processing the next hidden layer.

\begin{figure*}
\centering   
\includegraphics[width=\linewidth]{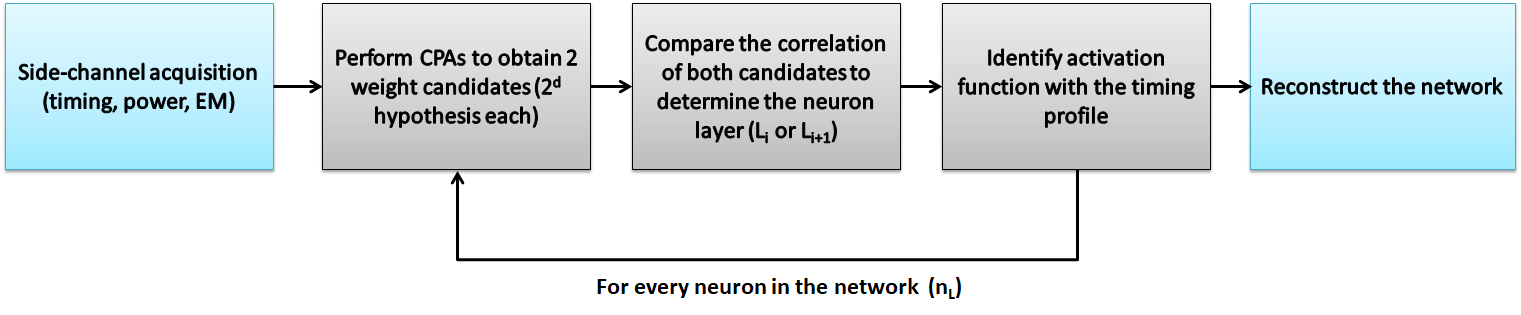}
	\caption{Methodology to reverse engineer the target neural network}
\label{procedure}
\end{figure*}

\subsection{Recovery of the Full Network Layout}
\label{sec:recover}
The combination of previously developed individual techniques can thereafter result in a full reverse engineering of the network.
The full  network recovery is performed layer by layer, and for each layer, the weights for each neuron have to be recovered one at a time.
Let us consider a network consisting of $N$ layers, $L_0, L_1, ..., L_{N-1}$, with $L_0$ being the input layer and $L_{N-1}$ being the output layer.
The reverse engineering is performed with the following steps:
\begin{enumerate}
\item The first step is to recover the weight $w_{L_0}$ of each connection from the input layer $(L_0)$ and the first hidden layer $(L_1)$. Since the dimension of the input layer is known, the CPA can be performed $n_{L_0}$ times (the size of $L_0$). The correlation is computed for $2^d$ hypotheses ($d$ is the number of bits in IEEE 754 representation, normally it is 32 bits, but to simplify, 16 bits can be used with lesser precision for the mantissa). After the weights have been recovered, the output of the sum of multiplication can be calculated. This information provides us with the input to the activation function
\item In order to determine the output of the sum of the multiplications, the number of neurons in the layer must be known. This can be recovered by the combination of SPA and DPA technique described in the previous subsection (2 times CPA for each weight candidate $w$, so in total $2n_{L_0}2^d$ CPA required), in parallel with the weight recovery. When all the weights of the first hidden layer are recovered, the following steps are executed.
\item Using the same set of traces, timing patterns for different inputs to the activation function can be built, similar to Figure~\ref{f:0}. Timing patterns or average timing can then be compared with the profile of each function to determine the activation function (a comparison can be based on simple statistical tools like correlation, distance metric, etc). Afterward, the output of the activation function can be computed, which provides the input to the next layer.
\item The same steps are repeated in the subsequent layers ($L_1, ..., L_{N-1}$, so in total at most $2Nn_L2^d$, where $n_L$ is $max(n_{L_0},...,n_{L_{N-1}})$) until the structure of the full network is recovered.
\end{enumerate}
The whole procedure is depicted in Figure~\ref{procedure}.
In general, it can be seen that the attack scales linearly with the size of the network.
Moreover, the same set of traces can be reused for various steps of the attack and attacking different layers, thus reducing measurement effort.

\section{Single Trace Input Recovery Attack on MLP}
\label{sec:input_recovery}

\begin{figure}
\centering   
\includegraphics[width=0.8\linewidth]{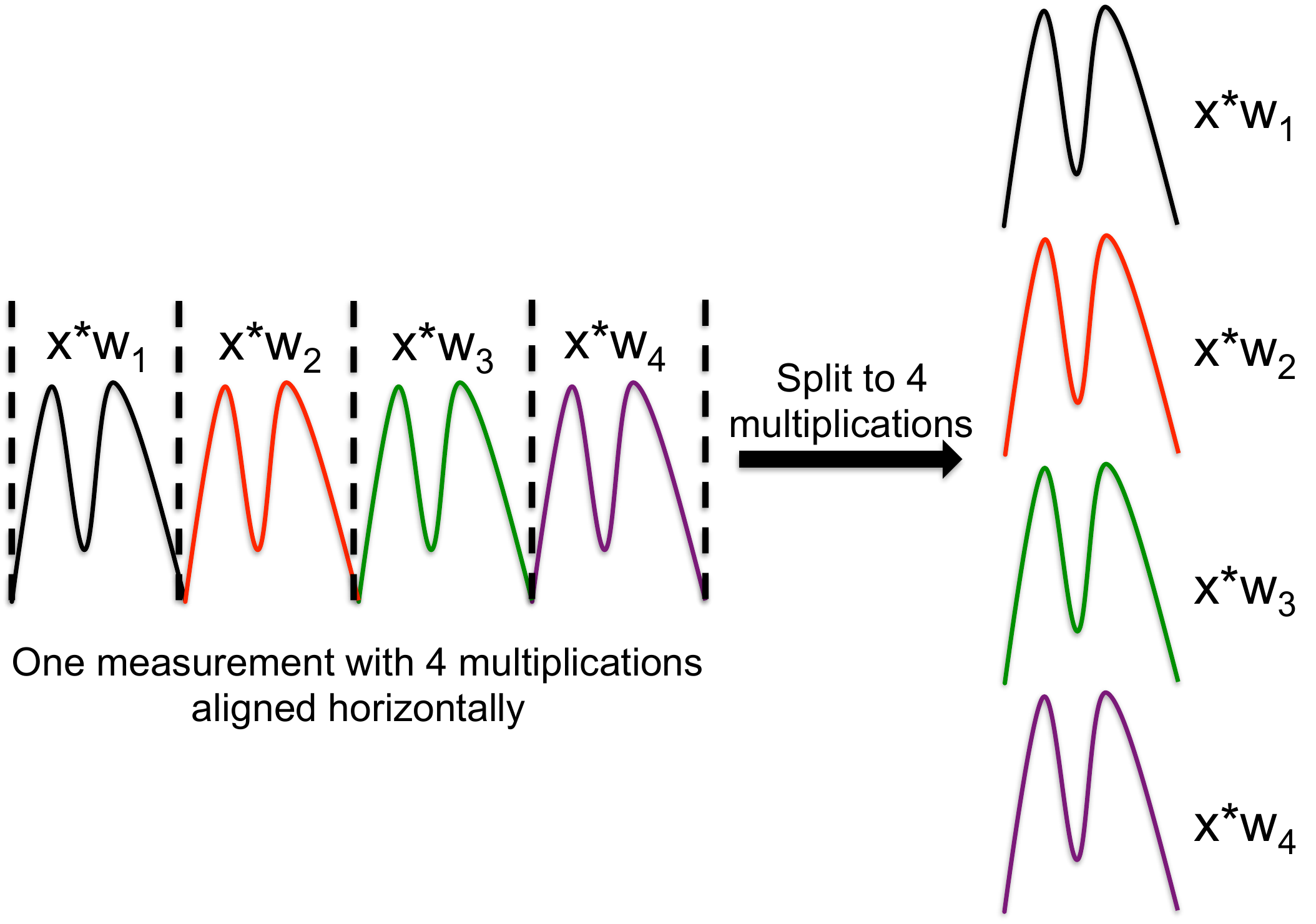}
	\caption{Illustration of a recovery of multiple measurements from a single measurement processing several elementary operations sequentially}
\label{hdpa}
\end{figure}

%\begin{figure*}
%\centering   
%	\subfloat[First byte mantissa of input]{\label{f:4a}\includegraphics[width=0.33\linewidth]{cpa2_r0_1}}
%\subfloat[Second byte mantissa of input]{\label{f:4b}\includegraphics[width=0.33\linewidth]{cpa2_r1_1}}
%\subfloat[Third byte mantissa of input]{\label{f:4c}\includegraphics[width=0.33\linewidth]{cpa2_r2_1}}
%\caption{Input recovery attack on the initial layer}
%\label{f:4}
%\end{figure*}

\begin{figure}
\centering   
\subfloat[First byte recovery (sign and 7-bit exponent)]{\label{f:4b}\includegraphics[width=0.8\linewidth]{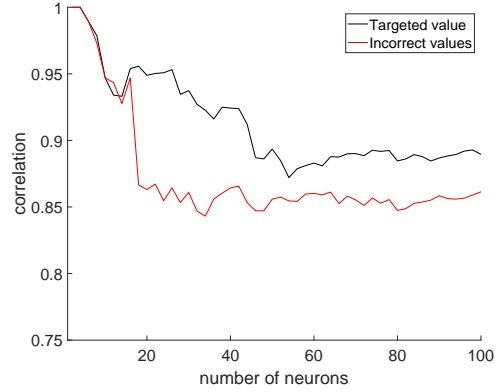}}\\
\subfloat[Second byte recovery (lsb exponent and mantissa)]{\label{f:4c}\includegraphics[width=0.8\linewidth]{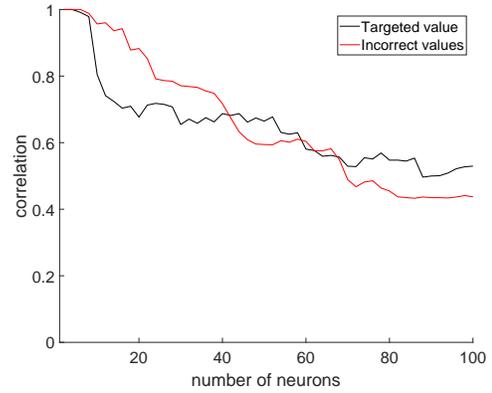}}
\caption{Input recovery attack on the initial layer}
\label{f:4}
\end{figure}

In the previous section, the methodology to reverse engineer a neural network has been described and practically demonstrated.
In this section, we consider an alternate scenario, where an unknown or secret input is fed to a known network.
By known network, we mean that the architecture and weights are either public or known to the adversary (e.g., recovered by reverse engineering).
Generally, it can be extremely complex to recover the input by observing outputs from a known network.
It involves several classifications in order to solve a system of equations, while some of the functions might not be invertible, i.e., ReLU.
When considering theoretical attacks, the system of equations can soon become unmanageable as the architecture of the network becomes complex.
%SCA can be applied to perform an input recovery attack.
%\todo{relates to what?}

The proposed attack targets the multiplication operation in the first hidden layer.
It is exactly the opposite of the previous weight recovery attack, as the weights $w$ are known while input $x$ is unknown.
However, there is a strong limitation with this attack.
As $x$ changes from one measurement to another, information learned from one measurement cannot be used with another measurement, preventing any statistical analysis. 
In this case, the adversary is forced to exploit all the measurements from a single measurement.
Thus, to perform information exploitation over a single measurement, we use HPA.
The weights in the first hidden layer are all multiplied with the same input $x$, one after the other.
Drawing analogy with SCA on cryptography, several known plaintexts (weights in case of MLP) are processed for a single unknown key (input $x$ here).
The only difference is that all the processing is done in different parts of a single trace.
An input recovery attack was proposed in~\cite{DBLP:journals/corr/abs-1803-05847}, which requires multiple traces targeting a line buffer, which is an optimization oriented design choice.
Contrary, our proposed attack targets the generic multiplication in a single trace setting.

We measured the EM trace to perform an input recovery attack.
$M$ multiplications, corresponding to $M$ different weights (or neurons), in the first hidden layer were isolated.
An illustrative example is shown in Figure~\ref{hdpa} where $M=4$ traces corresponding to 4 weights are recovered from a single trace.
Thus, a single trace is cut into $M$ smaller traces, each one corresponding to one multiplication with an associated weight.
Next, the value of the input is statistically inferred by applying a standard DPA on the $M$ smaller traces.
The results are shown in Figure~\ref{f:4} for different bytes of the same input.
The black curve shows the correlation of the correct input while all wrong inputs are represented in red.
The attack needs 20 or more multiplications to reliably recover the input.
This means that in the current setting, the proposed attack works very well on medium to large sized networks, with at least 40 neurons in the first hidden layer (which is no issue in modern architectures used today).

%Once the reverse engineering of the neural network has been done, the layout can be used to perform secret input recovery. 
%One idea is to use the classification output and layout of the network, however, it will require several classifications in order to solve a system of equations, not to mention some of the functions might not be invertible, i.e., ReLU.
%Instead, the more suitable target is the multiplication operation in the first hidden layer. In this scenario, it will be similar to weight recovery phase, except the role will be swapped, here we know the weights $w$, but not the input value $x$.
%
%One of the potential limitation is that, unlike input value which can be collected as many as we want, the weights are limited by the number of neurons in the first hidden layer. We conducted the experiments by varying number of neurons in the first hidden layer, to increase the number of "traces" for the attack. In Figure~\ref{},\textcolor{red}{[Add the bullet points here]}

\section{Experimental Validation on ARM Cortex-M3}
\label{sec:arm}
A methodology to reverse engineer sensitive parameters of a neural network and input recovery was proposed in previous sections.
The attack was practically validated on an 8-bit AVR (Atmel ATmega328P).
In this section, we extend the presented attack on a 32-bit ARM microcontrollers.
ARM microcontrollers form a fair share of the current market with huge dominance in mobile applications, but also seeing rapid adoption in markets like IoT, automotive, virtual and augmented reality etc.

Our target platform is the widely available \emph{Arduino due} development board which contains an \emph{Atmel SAM3X8E ARM Cortex-M3} CPU with a 3-stage pipeline, operating at 84 MHz. The measurement setup is similar to previous experiments (Lecroy WaveRunner 610zi oscilloscope and RF-U 5-2 near-field EM probe from Langer).
The point of measurements was determined by a benchmarking code running AES encryption.
After capturing the measurements for the target neural network, one can perform the reverse engineering. % by a step by step approach.

\begin{figure}
\centering   
\subfloat[ReLU]{\label{arm:0a}\includegraphics[width=0.5\linewidth]{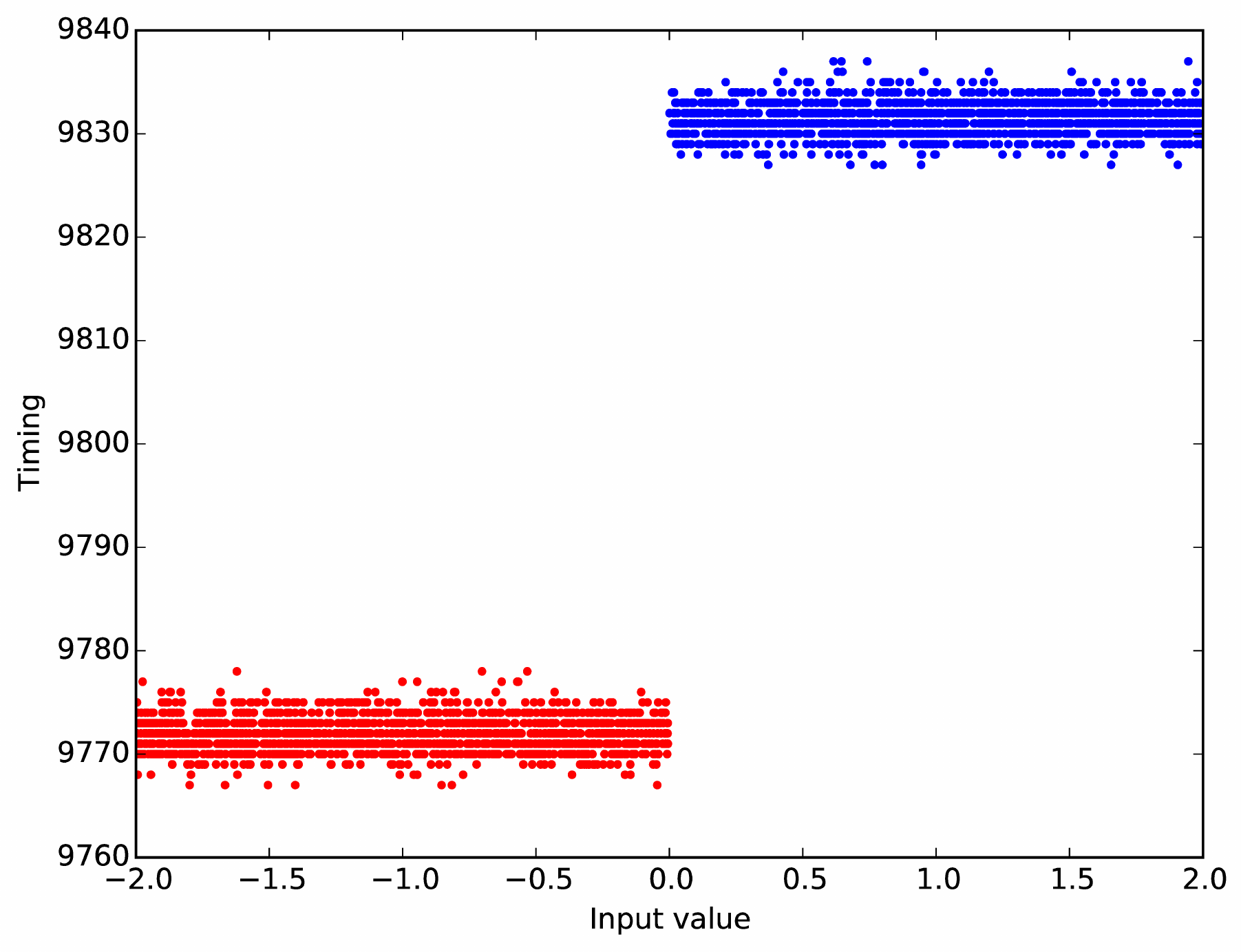}}
\subfloat[Sigmoid]{\label{arm:0b}\includegraphics[width=0.5\linewidth]{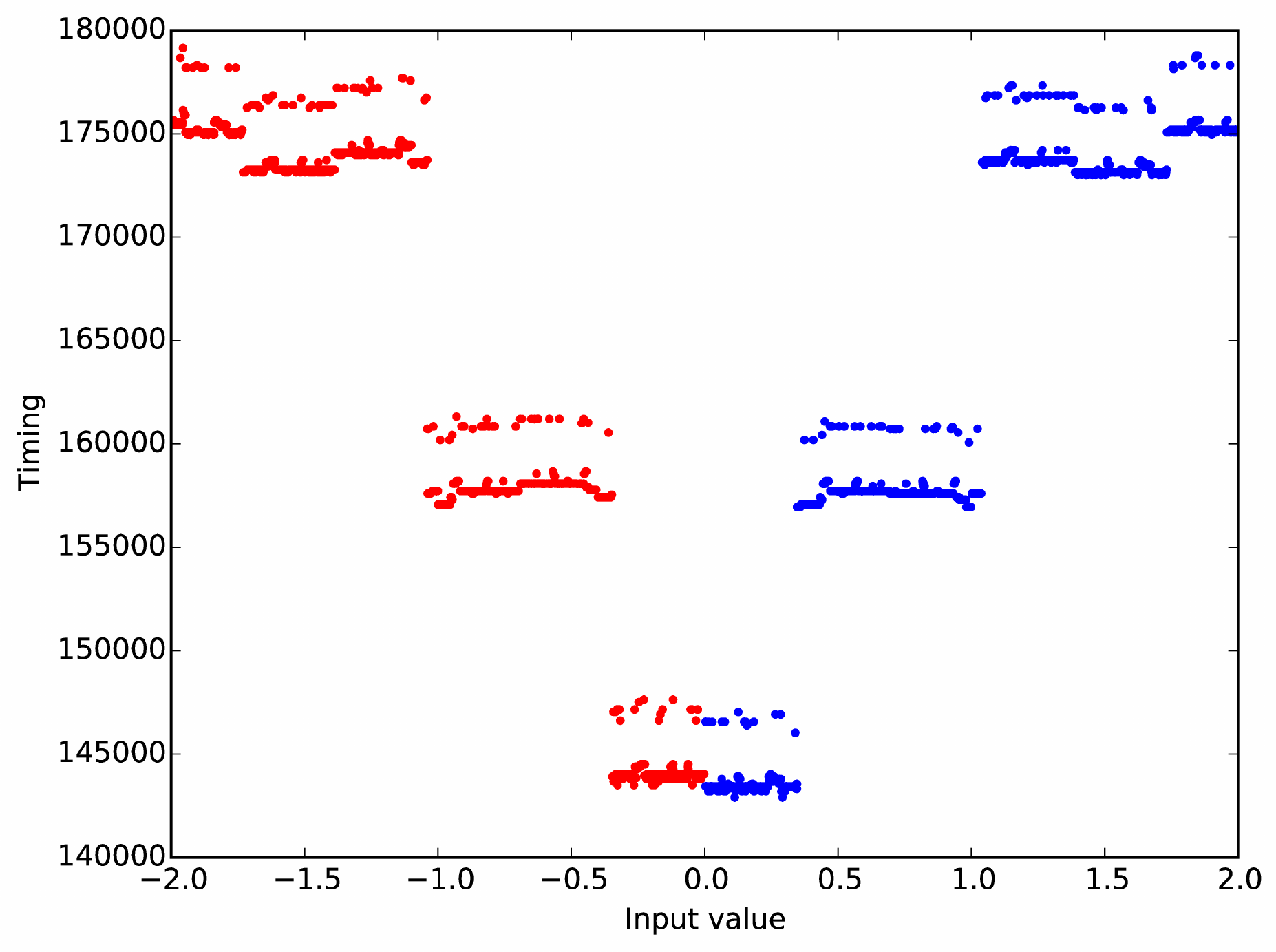}}\\
\subfloat[Tanh]{\label{arm:0c}\includegraphics[width=0.5\linewidth]{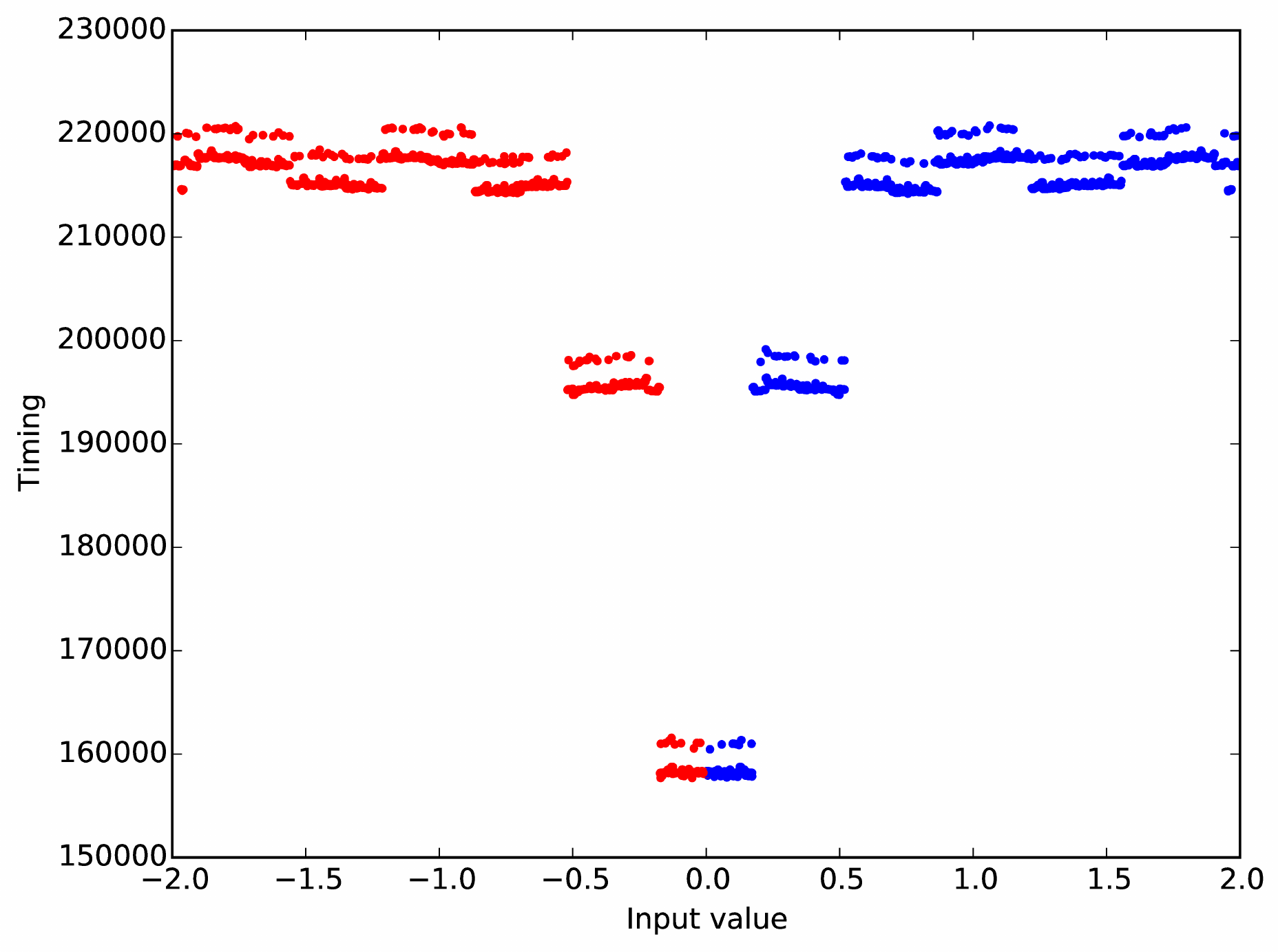}}
\caption{Timing behavior for different activation functions}
\label{timing-arm}
\end{figure}

\begin{figure}
\centering   
\includegraphics[width=0.8\linewidth]{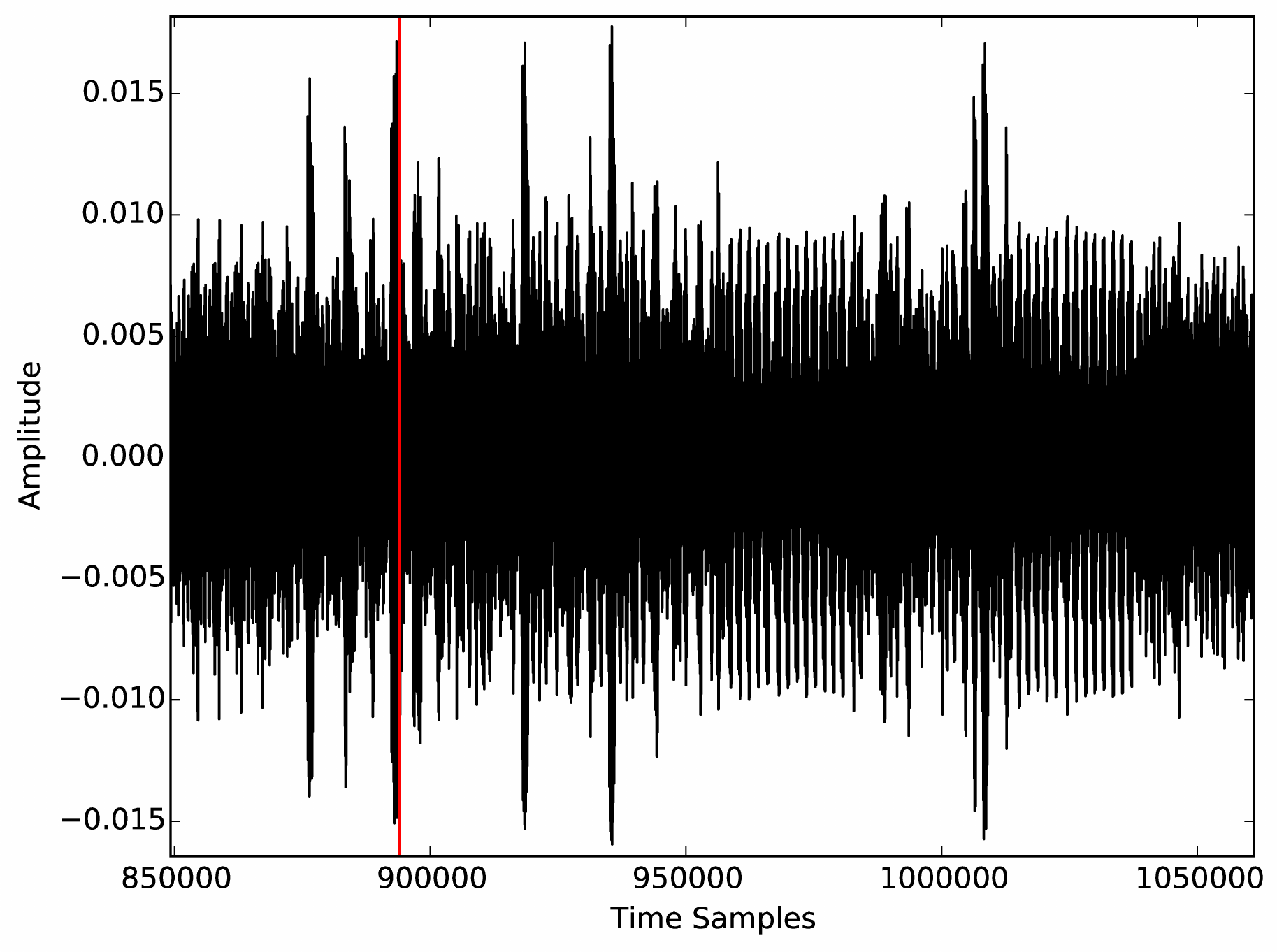}
	\caption{Observing pattern and timing of multiplication and activation function}
\label{onel-arm}
\end{figure}

The timing behavior of various activation functions are shown in Figure~\ref{timing-arm}.
The results, though different from previous experiments on AVR, have unique timing signatures, allowing identification of each activation function.
The activity of a single neuron is shown in Figure~\ref{onel-arm}, which uses sigmoid as an activation function (separated by multiplication a vertical red line). 
%By knowing the timing behavior of the activation function, we can perform a SPA to identify multiplication and activation operation in the captured measured (shown separated by a vertical red line).

\begin{figure}
\centering   
\includegraphics[width=0.8\linewidth]{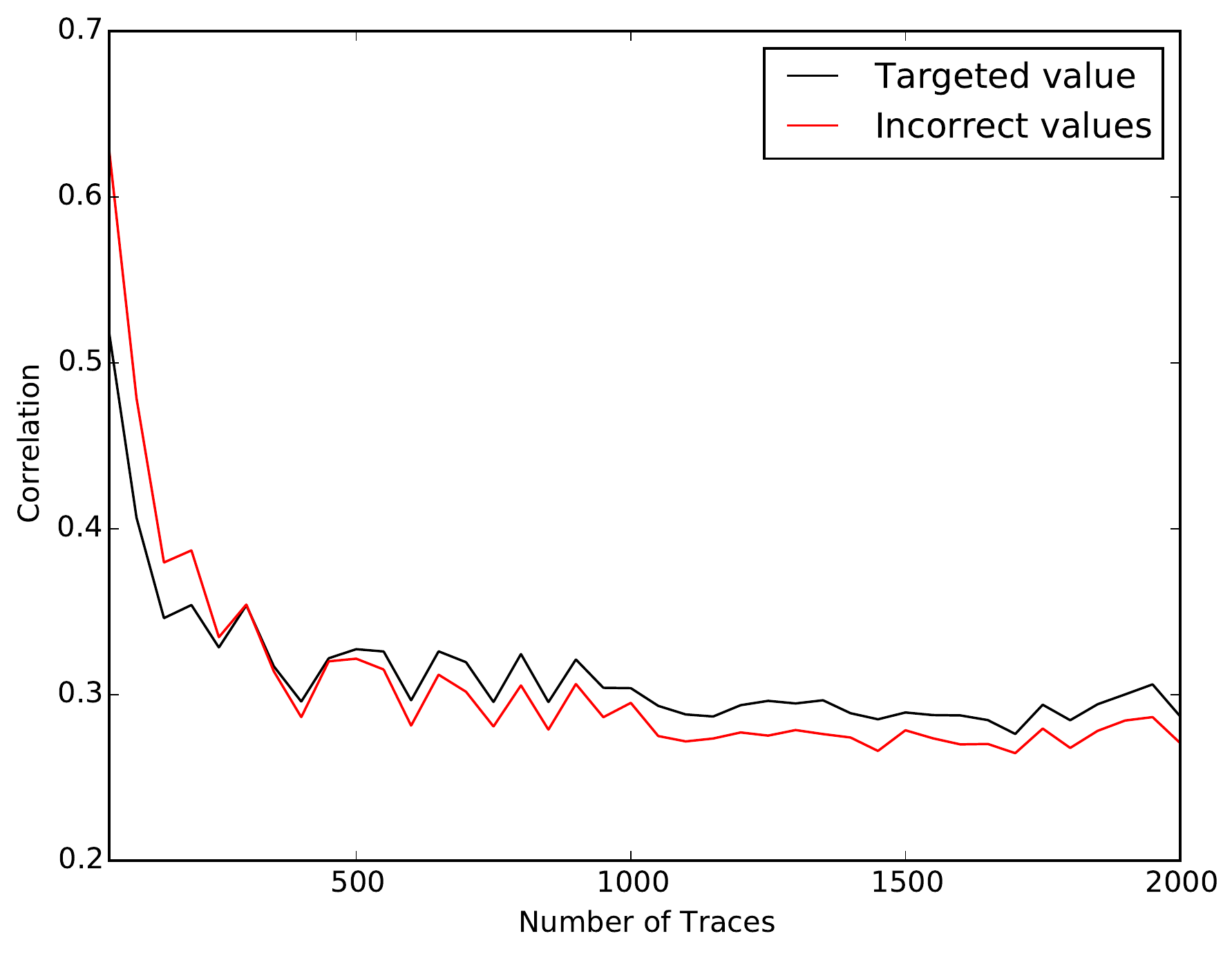}
	\caption{Correlation comparison between correct and incorrect mantissa for weight=$2.453$}
\label{mult-arm}
\end{figure}

A known input attack is mounted on the multiplication to recover the secret weight.
One practical consideration in attacking multiplication is that different compilers will compile it differently for different targets.
Modern microcontrollers also have dedicated floating point units for handling operations like multiplication of real numbers.
To avoid the discrepancy of a difference of multiplication operation, we target the output of multiplication.
In other words, we target the point when multiplication operation with secret weight is completed and the resultant product is updated in general purpose registers or memory.
Figure~\ref{mult-arm} shows the success of attack recovering secret weight of $2.453$, with known input.
As stated before, side-channel measurements on modern 32-bit ARM Cortex-M3 may have lower SNR thus making attack slightly harder.
Nevertheless, the attack is shown practical even on ARM with $2\times$ more measurements.
In our setup, getting $200$ extra measurement takes less than a minute.
Similarly, the setup and number of measurements can be updated for other targets like FPGA, GPU, etc.

\begin{figure}
\centering   
\includegraphics[width=0.8\linewidth]{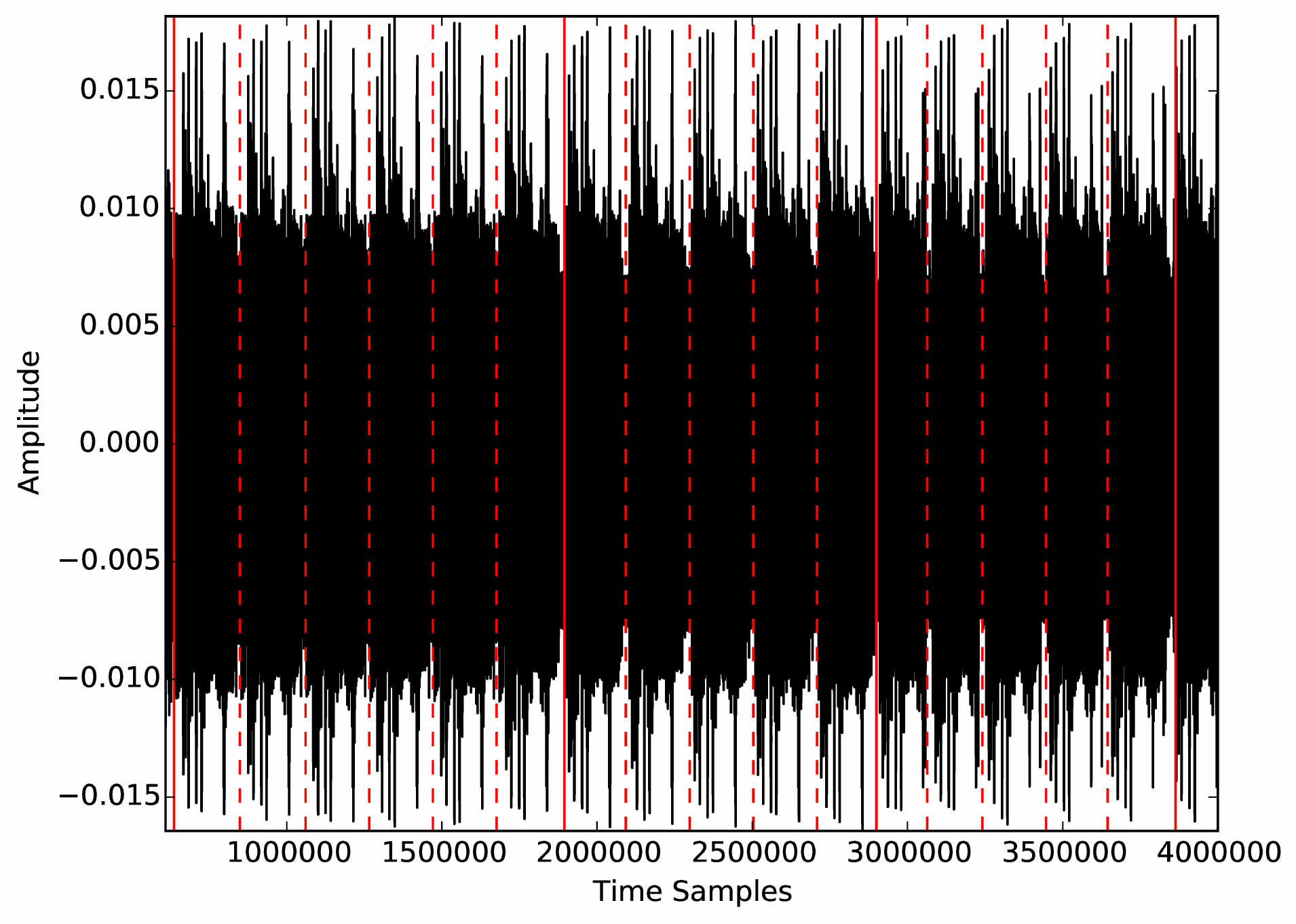}
\caption{SPA on hidden layers with 3 hidden layers (6,5,5 neurons each)}
\label{full-arm}
\end{figure}

Finally, the full network layout is recovered.
The activity of a full network with 3 hidden layers composed of 6, 5, and 5 neurons each is shown in Figure~\ref{full-arm}.
All the neurons are observable by a visual inspection.
The determination of layer boundaries (shown by solid red line) can be determined by attacking the multiplication operation and following the approach discussed in Section~\ref{sec:recover}.

\section{Mitigations}
\label{sec:countermeasures}

As demonstrated above, various side-channel attacks can be applied to reverse engineer certain components of a pre-trained network.
To mitigate such a recovery, several countermeasures can be deployed:
\begin{enumerate}
	\item Hidden layers of an MLP must be executed in sequence but the multiplication operation in individual neurons within a layer can be executed independently. An example is shuffling~\cite{veyrat2012shuffling} as a well-studied side-channel countermeasure. It involves shuffling/permuting the order of execution of independent sub-operations. For example, given $N$ sub-operations ($1,\ldots,N$) and a random permutation $\sigma$, the order of execution becomes $(\sigma (1),\ldots, \sigma (N))$ instead. In this case, we propose to shuffle the order of multiplications of individual neurons within a hidden layer during every classification step. Shuffling modifies the time window of operations from one execution to another, mitigating a classical DPA attack.
	\item Weight recovery, as well as the single trace input recovery, can benefit from the application of masking countermeasures~\cite{coron2000boolean,prouff2013masking}. Masking is another widely studied side-channel countermeasure that is even accompanied by a formal proof of security. It involves mixing of sensitive computations with random numbers to remove the dependencies between actual data and side-channel signature, thus preventing the attack. For every operation $f(x,w)$, it is transformed into $f_m(x \oplus m_1, w \oplus m_2) = f(x,w) \oplus m$, where $m, m_1, m_2$ are uniformly drawn random mask, and $f_m$ is the masked function which apply mask $m$ at the output of $f$, given masked inputs $x \oplus m_1$ and  $w \oplus m_2$. If each neuron is individually masked with an independently drawn uniformly random mask for every iteration and every neuron, the proposed attacks can be prevented. However, this might result in a substantial performance penalty.
	\item The proposed attack on activation functions is possible due to the non-constant timing behavior. Mostly considered activation functions perform exponentiation operation. Implementation of constant time exponentiation has been widely studied in the domain of public key cryptography~\cite{hachez2000montgomery}. These well-studied ideas can be adjusted to implement constant time activation function processing.
\end{enumerate}

Clearly, all those countermeasures come with an area and performance cost.
In particular, shuffling and masking require a true random number generator that is typically very expensive in terms of area and performance.
%The computation with masking in a secure manner can also result in several fold area increase or performance degradation.
Similarly, constant time implementations of exponentiation~\cite{al2008comparative} also come at performance efficiency degradation.
Thus, the optimal choice of protection mechanism should be done after a systematic resource and performance evaluation study.

\section{Further Discussions and Conclusions}
%\label{sec:discussion}
%\section{Conclusion}
\label{sec:conclusion}

%Conclusion
Neural networks are widely used machine learning family of algorithms due to its versatility across domains.
Their effectiveness depends on the chosen architecture and fine-tuned parameters along with the trained weights, which can be a proprietary information.
In this work, we practically demonstrate reverse engineering of a neural network using side-channel analysis techniques.
Practical attacks are performed on measured data corresponding to chosen networks. To make our setting more general, we do not assume any specific form of the input data (except that inputs are real values).

We conclude that using an appropriate combination of SPA and DPA techniques, all sensitive parameters of the network can be recovered.
Moreover, a powerful HPA method is used to recover secret inputs from a known network in a single shot side-channel analysis.
The proposed methodology is practically demonstrated on two different modern controllers, a classic 8-bit AVR and a modern 32-bit ARM Cortex-M3 microcontroller.
As shown, the attack on modern devices are slightly harder to mount due to lower SNR for side-channel attacks but are still practical.
In the presented experiments, the attack took $2\times$ extra measurement, which require roughly $20$ seconds extra measurement time.
Overall, the attack methodology scales linearly with the size of the network.
%Our experiments are performed on MLP architectures that are implemented on an 8-bit AVR microcontroller.
%%Although these points could be considered to be drawbacks in our approach, we do not consider them to be so.
%Using this platform is indeed an option that is rather less encountered in practice (e.g., industry) but there are no elements of our attacks that could be considered as limiting factors for other targets like GPUs or FPGAs. The attacker might need more measurements or other enhanced signal processing methods to reach the same performance level as we demonstrate here, but there is no limitation in collecting the measurements from other platforms and reverse engineer the networks.

Multilayer perceptron architectures are widely used but arguably not the most common choice in state-of-the-art applications. Modern deep learning techniques like convolutional neural networks or recurrent neural networks recently took over and judging on the results they will remain as preferred methods of choice in coming years.  Yet, even those networks use the same activation functions we consider here as well as the fully connected layers (the difference is that they also have other types of layers). Since we are able to differentiate between the same type of layers in architectures, we expect the difference to be even more profound when comparing with other layer types.

When considering the weight vectors, here we consider the case where each node has a separate weight. Convolutional neural networks can actually also share those weights to lower the degree of the problem. The same technique we use here to obtain the independent weights can be used to obtain the shared weights (with in the worst case scenario, multiple unnecessary calculations for those shared weights).

The proposed attacks are both generic in nature and more powerful than the two previous works in this direction.
Finally, suggestions on countermeasures are provided to help designer mitigate such threats.
However, the proposed countermeasures are borrowed mainly from side-channel literature and can incur huge overheads. Nevertheless, we believe that they could motivate further research on optimized and effective countermeasures for neural networks.
Besides continuing working on countermeasures, as the main future research goal we envision the need to explore other types of layers, like convolution layers or max pooling layers. 

\balance

\bibliographystyle{IEEEtran}
\bibliography{bibliography}

\end{document}